\documentclass{aastex631}

\usepackage{amsmath}

\newcommand{\lyal}{Lyman-$\alpha$ }
\newcommand{\lyals}{Lyman-$\alpha$}
\newcommand{\hi}{\ion{H}{1}}

\begin{document}

\title[\lyal photons during CD]{Thermal Evolution of the Intergalactic Medium due to \lyal photons during the Cosmic Dawn}

\author[0000-0001-7451-6139]{Janakee Raste}
\affiliation{National Centre for Radio Astrophysics, 
Tata Institute of Fundamental Research, Pune 411007, India}

\author{Anjan Kumar Sarkar}
\affiliation{National Centre for Radio Astrophysics, 
Tata Institute of Fundamental Research, Pune 411007, India}

\author{Shiv K. Sethi}
\affiliation{Raman Research Institute, Bangalore 560080, India}

\correspondingauthor{Janakee Raste}
\email{janakee@ncra.tifr.res.in}

\begin{abstract}
The first star-forming objects that formed at high redshifts during the cosmic dawn (CD) also emitted photons between Lyman-$\alpha$ and Lyman-limit frequencies. These photons are instrumental in coupling the spin temperature of the neutral hydrogen (\hi) atoms with the kinetic temperature of the intergalactic medium (IGM). Along with this coupling effect, these photons also impact the kinetic temperature by exchanging energy with the \hi\ atoms. The injected Lyman-$\alpha$ photons in general cool the medium, while the continuum photons heat the medium. While studying this effect in the literature, quasi-static profile around the Lyman-$\alpha$ frequency is assumed. In this paper, we solve the time-dependent coupled dynamics of the photon intensity profile along with the evolution of the thermal state of the IGM and \hi\ spin temperature. It is expected that, during the CD era, the IGM has a mix of continuum photons with 10--20\% of injected photons. For this case, we show that the system reaches thermal equilibrium in around 1~Myr, with the final temperature in the range 50--100 K. This time scale is comparable to the source lifetime of PopIII stars at high redshifts. One impact of switching off short-lived sources is that it can keep the system heated above the temperature of the quasi-static state. We also show that the quasi-static equilibrium for the continuum photons is only achieved on time scales of 100~Myr at $z\simeq 20$, comparable to the age of the Universe. We also briefly discuss how the Lyman-$\alpha$ induced heating can impact the 21~cm signal from CD.
\end{abstract}

\keywords{
\href{http://astrothesaurus.org/uat/1285}{Population III stars (1285)}; 
\href{http://astrothesaurus.org/uat/1383}{Reionization (1383)}; 
\href{http://astrothesaurus.org/uat/813}{Intergalactic medium (813)}; 
\href{http://astrothesaurus.org/uat/690}{H I line emission (690)}; 
\href{http://astrothesaurus.org/uat/343}{Cosmology (343)}
}

\section{Introduction} \label{sec:intro}

Based on the $\Lambda$CDM model, our current understanding of the history of the Universe suggests that the dark ages of the Universe ended around redshift $z \simeq 35$ with the formation of first large-scale structures (the era of cosmic dawn (CD)). These subgalactic  structures 
emitted radiation,  which  heated and ionized the surrounding medium, a process culminating in the ionization of the Universe by $z \simeq 6$  (Epoch of reionization (EoR)) \citep{Barkana:2000fd,21cm_21cen,2014PTEP.2014fB112N,2010ARA&A..48..127M}. The physics of the first stars and galaxies is only partially understood theoretically and poorly constrained with observations \citep{yan2022first, hassan2023jwst}. 

The detection of the EoR/CD eras remains an outstanding aim of modern cosmology. The CMB temperature and polarization anisotropy detections by WMAP and Planck give partial information about these eras and determine the  redshift of reionization,  $z_{\rm reion} = 7.75 \pm 0.73$ \citep{Hinshaw:2012aka,Planck2018}.
These observations, along with Gunn-Peterson test on high-redshift quasars \citep{Fan:2000gq,2001AJ....122.2850B}, suggest that the Universe made a transition from nearly fully neutral to highly ionized in the 
redshift range $6 <z < 10$. Recent observations of \lyal forest suggest that the EoR ended around $z\simeq 5.3$ \citep{2023ApJ...955..115Z, 2021MNRAS.508.1853B, 2022MNRAS.514...55B, 2019MNRAS.485L..24K, RK21}.

The most direct   probe of  the era of EoR and CD  is through the detection of redshifted hyperfine 21~cm line of neutral hydrogen (\ion{H}{1}). This signal carries crucial information about the first sources of radiation in the Universe  \citep{Sethi05, 2006PhR...433..181F}. In particular, the \hi\ signal  from CD/EoR probes  the spectrum of sources in three frequency bands: hydrogen-ionizing radiation, radiation between Lyman-$\alpha$ frequency and Lyman-limit (referred to as `Lyman-$\alpha$ radiation, this radiation  determines the relative population of neutral hydrogen atoms in hyperfine states), and X-ray photons (which heat and partially ionize the medium) \citep{RS18, RS19}. In addition, the sources that  emit soft radio photons  also can impact  the observable \ion{H}{1} signal.

In this paper, we investigate the role of Lyman-$\alpha$ radiation in heating/cooling the neutral medium  during the epoch of cosmic dawn
(see e.g. \cite{mittal2023radiative,munoz2022impact,ghara2020impact,semelin2023accurate,meiksin2021impact,shimabukuro2023exploring,mcquinn2016deciphering} and references therein). These photons are characterized by  the spectral shape of their sources: continuum photons arise from stellar sources and 
correspond to flat spectrum between Lyman-$\alpha$ and Lyman-limit frequencies, and injected photons are generated by the absorption  of continuum photons to    higher atomic levels of  neutral hydrogen ($n\ge 3$).  
After an initial claim that Lyman-$\alpha$ radiation can cause substantial heating of the neutral hydrogen  \citep{1997ApJ...475..429M}, \citet{2004ApJ...602....1C} (CM04) showed that the heating/cooling of the medium is negligible for 
the number density of Lyman-$\alpha$ photons needed to couple the \hi\ hyperfine levels  to Lyman-$\alpha$ photons (Wouthuysen-Field effect). The results of CM04 have since been confirmed by other authors (e.g. \cite{chuzhoy2006ultraviolet,chuzhoy2007heating,rybicki2006improved,meiksin2006energy,hirata2006wouthuysen}). \citet{chuzhoy2006ultraviolet} (CS06) and \citet{chuzhoy2007heating} derived relations between
cooling/heating  rates owing to injected/continuum photons. Based on the  balance between heating and cooling, they argued  that the temperature in the medium could rise to around 200~K at $z \simeq 20$ (see also \cite{ciardi2010lyalpha,basko1981thermalization,chugai1987scattering,mittal2021ly,venumadhav2018heating} for more recent works and a  broader perspective on the issue). These works assume steady state conditions
for the energy exchange between photons and hydrogen atoms.  This assumption  is partly justified if all the relevant time scales 
exceed the time scale of reaching the quasi-steady state.   However, this doesn't allow for  a  self-consistent approach to
equilibrium of both the  photon distribution and  the thermal state of the medium.  In this paper, we extend this analysis
to solve the most-general time-dependent set of equations. One aim of our work is to seek theoretical completeness on this issue.

As discussed in detail in \cite{1994ApJ...427..603R} (RD94), the time scale at which
the steady state is reached could be  of the order of a few times the scattering time, $t_{\rm sca}$, multiplied by the Gunn-Peterson optical
depth $\tau_{\rm GP}$ (to be defined later).  This suggests that only sources which have life spans shorter than this time scale
could violate the steady state assumption. Based on our understanding of the ionizing sources, we expect both continuum  and injected  photons in the intergalactic medium (IGM) at the era of CD (e.g. CS06 and references therein).  Our analysis shows that for injected photons, the time scale at which the steady state is reached
is shorter than 1 million yr and is comparable to  $t_{\rm sca} \tau_{\rm GP}$. However, the time scale of reaching the quasi-steady state is much longer than this  expectation  for continuum photons,  and could be comparable to the expansion time scale. 

This provides further motivation for  us to  study the time-dependent heating/cooling of the IGM owing to the interaction of Lyman-$\alpha$ photons with the 
neutral medium. Our study  generalizes the analysis of RD94 with the inclusion of the  recoil term.  In our theoretical study, we also  rederive the energy exchange owing to Compton-inverse Compton scattering
between photons and atoms and generalize the expression derived by CS06. In the next section,
we discuss the formalism in detail and derive the general time-dependent expressions for relevant quantities. We also consider the steady-state limit of the time-dependent solutions, and explicitly 
demonstrate the equivalence of CM04 and CS06 approaches. The  main results are presented  in section~\ref{sec:resl}. In section~\ref{sec:conc}
we summarize our results and provide possible future avenues of our work.  Throughout the paper, we assume the best-fit Planck
parameters  corresponding to spatially flat FRW universe: $\Omega_c h^2 = 0.12$ and $\Omega_b h^2 = 0.0224$ \citep{Planck2018}.  

\section{Interaction of \texorpdfstring{Lyman-$\alpha$}{Lyman-alpha} photons with neutral hydrogen } \label{sec:lyah1}
Let $J(\hat n,\nu,{\bf x},t) \equiv dN/(d\nu dV d\Omega)$ denote the number density of photons of frequencies
between $\nu$ and $\nu+d\nu$, at a location ${\bf x}$, and  traveling in
a direction $\hat n$ (we have  mostly used notations from \cite{1994ApJ...427..603R} (RD94) and  \cite{2004ApJ...602....1C} (CM04) but we define $J$ as photon number density and not photon intensity).   For the FRW  universe, the relevant equation to second order in 
energy exchange $h\langle \Delta\nu \rangle$  is (e.g. CS06): 
\begin{equation}
{\partial J \over \partial t} - \nu {\dot a \over a} {\partial J \over d\nu} + 2{\dot a \over a} J  = n c \left [{\partial^2 (J\sigma) \over \partial \nu^2} \left \langle \left ({\Delta \nu\over 2} \right )^2\right \rangle - {\partial \over \partial \nu}(J\sigma) \langle \Delta\nu \rangle \right ] + C \psi(\nu).
\label{eq:eqtd}
\end{equation}
This equation generalizes the expression used by  CS06 for an expanding universe 
and nonzero source term. Here $\int \psi(\nu)d\nu = 1$  and $\psi$ is  normalized probability density.  $n$ is the number density of \hi\ atoms, $\sigma(\nu)= (3/8\pi)\lambda_\alpha^2 A_\alpha \phi(\nu)$ is the cross-section of scattering and other  variables have their usual meaning; $\phi(\nu)$ is the normalized response of the atom and is taken to  be Voigt profile in this work.   $C$ is rate of production  of new photons.
Eq.~(\ref{eq:eqtd}) can be further expressed  as
\begin{equation}
  {\partial J \over \partial t}  + 3{\dot a \over a} J  = {\partial \over \partial \nu} \left [H\nu J + nc{\partial (J\sigma) \over \partial \nu} { \langle \Delta \nu^2 \rangle \over 2}- nc(J\sigma) \langle \Delta\nu \rangle \right ] + C \psi(\nu).
  \label{eq:fineq}
\end{equation}
This form allows us to establish conservation of the number of  photons. For no injection of photons, or for  $C = 0$, Eq.~(\ref{eq:fineq}) 
integrated over all $\nu$  demonstrates  photon conservation in an expanding universe.  For our work, we define $C$ as the number density of new photons that are produced per unit time per unit solid angle    in the frequency range from Lyman-$\alpha$ to Lyman-limit (in $ \rm cm^{-3} \, sec^{-1} \, sr^{-1}$). 

There are two modes of energy transfer between  atoms and the radiation field:
(a) Compton and inverse Compton scattering, close to the Lyman-$\alpha$ resonant line, which determine the kinetic state of the
atom ($\Delta E_1$) and (b) Wouthuysen-Field  effect in which the energy exchange occurs owing to the change in  the level populations of hyperfine states  of neutral hydrogen ($\Delta E_2$). It can be shown that
\begin{eqnarray}
  \langle \Delta E_1 \rangle   & = &  {(h\nu_\alpha)^2 \over m_pc^2}\left (1 - kT_{\rm K} {\sigma'(\nu) \over \sigma(\nu)h} \right ) \quad {\rm and} \nonumber\\
  \langle \Delta E_2  \rangle  & = &  {b(h\nu_{21})^2 \over 2k T_{\rm S}}\left (1 - kT_{\rm S} {\sigma'(\nu) \over \sigma(\nu)h} \right ),
  \label{eq:ene_exchange}
\end{eqnarray}
with $b= 2/9$ and $h\langle \Delta \nu \rangle = \langle \Delta E_1 \rangle +\langle \Delta E_2 \rangle $ (for details of the computation of 
$\langle (\Delta \nu)^2 \rangle$, see CS06).  Here $\sigma'(\nu) = d\sigma/d\nu$. In the Appendix~\ref{app:eneex}, we rederive the energy exchange owing to Compton-inverse Compton scattering and show that there could be an additional term in the energy exchange, which is negligible for 
the homogeneous and isotropic IGM. 
Here, the spin temperature $T_{\rm S}$ is determined from 
\begin{equation}
  T_{\rm S} = {T_{\rm CMB} + y_c T_{\rm K} + y_\alpha T_\alpha \over 1+ y_c + y_\alpha},
  \label{eq:ts_def}
\end{equation}
and the color temperature $T_\alpha$, close to the resonance line is
\begin{align}
    T_\alpha = - {h(\nu - \nu_\alpha) \over k\; \; {\rm ln}\left( J(\nu) \over J(\nu_\alpha) \right)}.
    \label{eq:tadef}
\end{align}
The \lyal coupling coefficient $y_\alpha = (h\nu_{21}/kT_\alpha)(P_{21}/A_{21})$, where the rate of de-excitation of the upper hyperfine level is $P_{21} = (4/27) P_\alpha$ and the rate of transition 
owing to Lyman-$\alpha$ absorption by the \hi\ ground state is $P_\alpha \simeq  4\pi \sigma_0 J(\nu_\alpha) c$
(see e.g. \cite{2010ARA&A..48..127M, field1958excitation} for more details).  In thermal  equilibrium between Lyman-$\alpha$ photons and the IGM, $T_\alpha = T_{\rm K}$. 
If Lyman-$\alpha$ constitutes the dominating process for determining the spin temperature, $y_\alpha \gg y_c$ and $y_\alpha \gg 1$, then we have $T_{\rm S} = T_\alpha$. 

By a redefinition of variables, Eq.~(\ref{eq:fineq}) can be written as
\begin{equation}
  {1\over a^3}{\partial (J(x,t) a^3) \over \partial t'}    = {\partial \over \partial x} \left [{\phi(x) \over 2} {\partial J(x,t) \over \partial x} +\left(\eta\phi(x) + \gamma \right) J(x,t) \right ] + C' \psi(x) .
  \label{eq:fineq1}
\end{equation}
Here, $t' = t/t_{\rm sca}$ with the scattering time at the line center, $t_{\rm sca} =  1/(n \sigma_0 c(1+w/T_{\rm K}))$, and $x= (\nu-\nu_\alpha)/\Delta\nu_D$;  the thermal width, $\Delta\nu_D = \nu_\alpha v_T/c$ with thermal velocity $v_T = \sqrt{2k T_{\rm K}/m_p}$.  The parameter $w$ captures the contribution of hyperfine energy exchange,  $w =  b \nu_{21}^2 m_p c^2/(2\nu_\alpha^2 k)$\footnote{$w \simeq 0.4 \, \rm K$ and as $T_{\rm K}, T_{\rm S} \gg w$, the impact of hyperfine energy exchange is negligible except at very low temperatures.}. The Lyman-$\alpha$ scattering cross section at the line center is  $\sigma_0 = (3/8\pi) (\lambda_\alpha^2 A_\alpha/\Delta\nu_D)$, and the Sobolev parameter $\gamma = \tau_{\rm GP}^{-1} (1+w/T_{\rm K})^{-1}$ with the Gunn-Peterson (GP) optical depth $\tau_{\rm GP} =\sigma_0 n v_T/H$ ; $\gamma$ 
compares the relative efficacy of scattering and expansion. The recoil parameter  $\eta = (1+w/T_{\rm S})(1+w/T_{\rm K})^{-1} (h\nu_\alpha/(2kT_{\rm K}m_pc^2)^{1/2})$ and the rescaled photon injection rate is $C' = C/(n\sigma_0 c (1+w/T_{\rm K}))$.  Notice that $J(x,t)dx = J(\nu,t)d\nu$ which implies that $J(x,t) = J(\nu,t) \Delta\nu_D$. Similarly, $\psi(x) = \psi(\nu) \Delta\nu_D$.  We further redefine, $J(x,t) \to a^3 J(x,t)$ and $C' \to C' a^3$, which converts both the  photon intensity and the photon production rate into comoving quantities.   This gives
\begin{equation}
  {\partial J(x,t) \over \partial t'}    = {\partial \over \partial x} \left [{\phi(x) \over 2} {\partial J(x,t) \over \partial x} +\left(\eta\phi(x) + \gamma \right) J(x,t) \right ] + C' \psi(x).
  \label{eq:fineq2}
\end{equation}
Eq.~(\ref{eq:fineq2}) is similar to the one  derived by RD94 except for the factor of $a^3$ instead of $a^2$ and that Eq.~(\ref{eq:fineq2}) also correctly accounts  for energy exchange owing to  hyperfine mixing.  It can readily be shown that $C't' = Ct$ is the photon number density integrated over   all frequencies, which is consistent with conservation of the number of photons.  

As shown in RD94,  approximate steady state solutions can be obtained analytically by neglecting scattering (including the recoil term) in
Eq.~(\ref{eq:fineq2}). This reduces  Eq.~(\ref{eq:fineq2})  to
\begin{equation}
{\partial J \over \partial t'}   =  \gamma {\partial J \over \partial x} +  C'(t)  \psi(x).
\end{equation}
For $J(x,0) = 0$, it yields a solution (see e.g. RD94 for details):
\begin{equation}
  J(x,t') = {C' \over \gamma} \left [ \Phi(x) - \Phi(x+\gamma t) \right ],
  \label{eq:solnosc}
\end{equation}
where $\Phi(x) = \int_x^\infty \psi(x') dx'$.

Let us first consider injected photons with the delta function profile. In this case,  $\psi(x) = \delta(x)$. Eq.~(\ref{eq:solnosc}) yields
\begin{equation}
  J(x,t') = \left\{
    \begin{array}{cl}
        0 &\quad \> \> {\rm for} \> \>  x +\gamma t' < 0  \\
        {C'\over \gamma} & \quad  \> \> {\rm for} \> \> x < 0 \> \> {\rm and} \> \> x+\gamma t' > 0 \\
        0 & \quad \> \> {\rm for} \> \> x >0. 
    \end{array} \right.
  \label{eq:intdel}
\end{equation}

For this case, the intensity profile  is confined to $x < 0$ and  moves to progressively more  negative 
values of $x$ as the time progresses. Our main aim is to find the steady state profile close to the resonance
$|x| \lesssim 1$. For $|x| \simeq 1$, the steady state is reached after $t' \simeq 1/\gamma$. For a given time,
the smallest frequency with nonzero number of photons is $-\gamma t'$.  Integrating Eq.~(\ref{eq:intdel}) from 
this frequency to $x = 0$, we get the total number of photons to be  $C' t'= Ct$, in accordance with the conservation of number of photons. 

We next consider the continuum case. We assume a flat profile between $x_{\rm min} \ll 0$ and $x_{\rm max} \gg 0$. In a more realistic 
case, $x_{\rm max}$ could correspond to Lyman-$\beta$ frequency.  In this case, 
\begin{equation}
\psi(x) = {1\over x_{\rm max} - x_{\rm min}} \> \> {\rm for} \> \> x_{\rm min} < x < x_{\rm max}.
\end{equation}
This gives us
\begin{equation}
  \Phi(x) = \left\{
    \begin{array}{cl}
        1 & \quad {\rm for} \> \>  x \le x_{\rm min} \nonumber \\
        {(x_{\rm max} -x) \over (x_{\rm max} - x_{\rm min})} & \quad {\rm for} \> \> x_{\rm min} \le x \le x_{\rm max} \\
        0 & \quad {\rm for} \> \>  x \ge   x_{\rm max},  
    \end{array} \right.  \quad {\rm and}
\end{equation}
\begin{equation}
  \Phi(x+\gamma t') = \left\{
    \begin{array}{cl}
        1 & \quad {\rm for} \> \>  x \le x_{\rm min} -\gamma t' \\
        {(x_{\rm max} -x -\gamma t') \over (x_{\rm max} - x_{\rm min})} & \quad {\rm for} \> \> x_{\rm min} -\gamma t'  \le  x \le x_{\rm max} -\gamma t' \\
        0 & \quad {\rm for} \> \>  x \ge   x_{\rm max} -\gamma t'  .
    \end{array} \right.
  \label{eq:phisolcon}
\end{equation}
From Eqs.~(\ref{eq:phisolcon}) and~(\ref{eq:solnosc})) it follows that
\begin{equation}
    J(x,t') = \left\{
    \begin{array}{cl}
        {C' \over \gamma}  {(x +\gamma t' - x_{\rm min}) \over (x_{\rm max} - x_{\rm min})}  & \quad {\rm for} \> \> x_{\rm min} -\gamma t'  \le  x \le x_{\rm min}  \\
        {C' t' \over (x_{\rm max} - x_{\rm min})} & \quad {\rm for} \> \> x_{\rm min} \le  x \le  x_{\rm max} -\gamma t' \\ 
        {C' \over \gamma} {(x_{\rm max} -x) \over (x_{\rm max} - x_{\rm min})} & \quad {\rm for} \> \> x_{\rm max} - \gamma t' \le x \le x_{\rm max} .
    \end{array} \right.
\label{eq:jsolcont}
\end{equation}
It can be verified that  $J(x,t')$ (Eq.~(\ref{eq:jsolcont})) integrated over all frequencies ($x_{\rm min} -\gamma t' \le x \le x_{\rm max}$) yields
$C' t'$, consistent with the conservation of the number of photons.  Eq.~(\ref{eq:jsolcont}) further shows that, for $x +\gamma t' > x_{\rm max}$, we get the steady-state solution. For $|x| \lesssim 1 $, the time at which the quasi-equilibrium is reached is $t' \simeq x_{\rm max}/\gamma$. As noted above, $x_{\rm max}$ can be computed using Lyman-$\beta$ frequency, which gives  $x_{\rm max} \simeq 10^5$ (for $T = 10 \, \rm K$);  this time is orders of magnitude larger than  for the  injected photons
and could be comparable to  the expansion time scales.

The evolution of $J(x)$ and the spin temperature, $T_{\rm S}$ depends on the thermal state of the gas given by the kinetic temperature $T_{\rm K}$. Its
evolution, for a neutral, monoatomic,  gas,  is given by
\begin{equation}
{dT_{\rm K} \over dt} = -2 H T_{\rm K} + {2\over 3} {\dot q \over n_b k}.
\label{eq:thermal}
  \end{equation}
Here, $\dot q = \dot Q n_b  $ is the rate at which the energy is injected  per unit volume ($\rm erg \, cm^{-3} \, sec^{-1}$). 
Using Eq.~(\ref{eq:ene_exchange}) one can compute the rate at which the energy is pumped per baryon as
\begin{equation}
    \dot Q = 4\pi c\int J(\nu,t) \sigma(\nu) (\langle \Delta E_1 \rangle + \langle \Delta E_2 \rangle) d\nu.
    \label{eq:ene_exchange3}
\end{equation}
Eqs.~(\ref{eq:fineq2}) and~(\ref{eq:thermal}) along with Eqs.~(\ref{eq:ene_exchange3}) and~(\ref{eq:ts_def}) allow one
to formulate the initial value problem of the simultaneous evolution of the photon spectrum along with the hyperfine 
and thermal state  of the atom. 

To gain further insights into  this set of equations, in particular the approach to equilibrium/steady state, one can integrate   Eq.~(\ref{eq:ene_exchange3})  by parts, after inserting Eq.~(\ref{eq:ene_exchange}) in it
\begin{eqnarray}
    \dot Q &=&  4\pi c\int  {(h\nu_\alpha)^2 \over m_p c^2} \sigma_0\phi(x)  \left(J(x,t) + {kT_{\rm K} \over \Delta \nu_D h} J'(x,t) \right) dx\nonumber \\
    &+&  4 \pi c\int  {b(h\nu_{21})^2 \over 2k T_{\rm S}} \sigma_0\phi(x)  \left(J(x,t) + {kT_{\rm S} \over \Delta \nu_D h} J'(x,t) \right) dx.
    \label{eq:ene_exchange1}
\end{eqnarray}
Eq.~(\ref{eq:ene_exchange1}) generalizes the energy injection expression derived by CM04 and CS06 and  is applicable to  any time-dependent photon intensity. To connect this expression to the existing 
literature, we first consider a  steady-state solution of Eq.~(\ref{eq:fineq2}) (CM04, CS06): 
\begin{equation}
    \phi(x) J'(x) + 2\left(\eta \phi(x) + \gamma \right) J(x) = 2\gamma J_0 \left[1-k_\alpha \Theta(x) \right ]. 
    \label{eq:eqdis}
\end{equation}
Here,  $k_\alpha = 0$ for continuum photons. $k_\alpha = 1$ for injected photons (with delta function profile at the line center\footnote{In this work, we assume Voigt profile for injected photons. The delta function profile is used here for illustration.}) for $x> 0$ and zero otherwise. $J_0$ is the photon intensity for $|x| \gg 1$.
$\Theta(x)$ is the Heaviside function; it is unity for $x > 0$ and vanishes otherwise.  

Using Eq.~(\ref{eq:eqdis}) in Eq.~(\ref{eq:ene_exchange1}), we get, for continuum photons, assuming $T_{\rm S} = T_{\rm K}$,
\begin{eqnarray}
    \dot Q & = & 4\pi c\int dx \left({(h\nu_\alpha)^2 \over m_p c^2} + {b(h\nu_{21})^2 \over k T_{\rm K}}\right)\sigma_0 \nonumber \\
    &\times& \left[\phi(x) J(x) - {kT_{\rm K} \over \Delta \nu_D h}\left(2\eta \phi(x) J(x) +2\gamma J(x) -2\gamma J_0 \right)\right]. 
    \label{eq:ene_exch7}
\end{eqnarray}
This further reduces to
\begin{equation}
    \dot Q  = 4 \pi c\int   \left({(h\nu_\alpha)^2 \over m_p c^2} + {b(h\nu_{21})^2 \over k T_{\rm K}}\right)  \sigma_0 {2kT  \gamma\over \Delta\nu_D h}\left(J_0 - J(x)\right) dx.
    \label{eq:ene_exch8}
\end{equation}
Notice that there is no exchange of energy if $\gamma = 0$ as steady-state condition corresponds to thermal equilibrium for a nonexpanding universe.
Neglecting the energy exchange owing to hyperfine mixing, this expression further simplifies to
\begin{equation}
    \dot Q \simeq   {4\pi h \nu_\alpha H \over n_{\rm HI} }\int    \left(J_0 - J(x)\right) dx.
    \label{eq:ene_exch9}
\end{equation}
This was derived by CM04. This also demonstrates the equivalence of the approaches of 
CM04 and CS06.

Two cases of interest for  understanding the equilibrium/steady state are as follows:

{\it (a) Steady state without expansion and photon pumping}: In this case, $H= C = 0$,
and $\partial J/\partial t = 0$. The solution  of Eq.~(\ref{eq:fineq2}) is
$J(\nu)  = J(\nu_\alpha)\exp(-2\eta x)$ and $2\eta x = h(\nu-\nu_\alpha)/k T_{\rm K}$\footnote{We expect the equilibrium profile around $\nu = \nu_\alpha$ to have a blackbody shape. There are two reasons why that is not the case. First, the occupation number of Lyman-$\alpha$ photons is too small for the stimulated emission to play any role, and therefore, it has not been included in Eq.~(\ref{eq:fineq2}). Second, $(\nu/\nu_\alpha)^2$ prefactors have also been dropped because they result in  energy exchange correction to an  order higher than two, which we do not require. In this limit, we have the correct equilibrium shape (for details, see \cite{rybicki2006improved}).}.  From Eq.~(\ref{eq:tadef}) this gives $T_\alpha = T_{\rm K}$. If
the density of  Lyman-alpha photons is large enough  such that $y_\alpha \gg y_c$ and $y_\alpha \gg 1$,  it further follows from Eq.~(\ref{eq:ts_def}), that $T_{\rm S} = T_\alpha = T_{\rm K}$.

For this case,  Eq.~(\ref{eq:ene_exchange1})  shows that the rate of energy exchange between photons and atoms can be written as 
\begin{equation}
    \dot Q = 4\pi c\int  \left ({(h\nu_\alpha)^2 \over m_p c^2} + {b(h\nu_{21})^2 \over k T_{\rm K}}\right) \sigma_0\phi(x)  \left(J(x) + {kT \over \Delta\nu_D  h} J'(x) \right) dx,
    \label{eq:ene_exch6}
\end{equation}
and it vanishes from Eq.~(\ref{eq:eqdis}).

{\it (b) Stationary state with expansion and photon pumping}: In this case,
there might not be  thermal equilibrium. However, a stationary photon profile  is possible
even without scattering as shown in RD94.
The addition of scattering including atom recoil  alters this stationary state close
to $x =0$. As  follows from Eq.~(\ref{eq:ene_exch8}), the energy exchange is proportional
to the expansion rate in this case. 

If $\partial J/\partial t' = 0$, then Eq.~(\ref{eq:fineq2}) can 
be expressed as
\begin{equation}
   {\partial \over \partial x} \left [{\phi(x) \over 2} {\partial J(x,t) \over \partial x} +\left(\eta\phi(x) + \gamma \right) J(x,t) \right ] + C' \psi(x) = 0.
  \label{eq:fineq2ss}
\end{equation}
The scattering impacts the photon spectrum for $x\simeq 0$ and therefore, the no-scattering solutions  (Eqs.~(\ref{eq:intdel}) and~(\ref{eq:jsolcont})) should  yield  correct  solutions  for  $|x| \gg 0$.  Eq.~(\ref{eq:fineq2ss}) is valid for the range 
of $x$ for which the intensity has reached a quasi-steady state. Eqs.~(\ref{eq:intdel}) and~(\ref{eq:jsolcont}) show that, 
for $x\ll 0$, the intensity continues to evolve even for large times, but it reaches steady state at earlier times  
for $x \gtrsim 0$. This motivates us to integrate Eq.~(\ref{eq:fineq2ss}) from $x$ to $x_{\rm max}$, where $x$ corresponds
to frequencies at which the Voigt profile $\phi(x)$ is still substantial  ($|x| \simeq 0$). As $\phi(x)$  fall rapidly for 
$|x| \gg 0$, this gives us
\begin{equation}
   \phi(x){\partial J(x) \over \partial x} +2\left(\eta\phi(x) + \gamma \right) J(x) - 2C' \int_x^{x_{\rm max}}\psi(x') = 0.
  \label{eq:fineq2ss1}
\end{equation}
Eq.~(\ref{eq:fineq2ss1}) generalizes Eq.~(\ref{eq:eqdis}) for $C' \ne 0$ and when the correct  boundary condition for continuum photons 
is used ($J(x) = 0$ for $x \ge x_{\rm max}$). We note that the two equations yield the 
same answer for injected photons as $J_0 = C'/\gamma$. The difference between the two cases is greater for continuum
photons as they are based on different equilibrium profiles. However, in both cases, the difference is proportional to 
$C' \simeq \gamma J(0)$. As noted above, for $C' = \gamma = 0$, the steady-state  yields the equilibrium solution  $J_{\rm eq} = J_{\rm eq}^0 \exp(-2\eta x)$, for which the heating rate vanishes. To simplify Eq.~(\ref{eq:fineq2ss1}) further, we note that the heating rate is
proportional to $\gamma \ll 1$, which motivates us to consider $J(x) = J_{\rm eq} + J_1$, where $J_1 \propto \gamma$ and terms second
order in $\gamma$ can be dropped. This allows to solve Eq.~(\ref{eq:fineq2ss1}) for an arbitrary $\phi(x)$:
\begin{equation}
   J(x) =  \exp(-2\eta x) \left[\int_0^x dx'  \left ( \left (2C' \exp(2\eta x')\int_{x'}^\infty dx'' \psi(x'')  - 2\gamma J_{\rm eq}^0 \right )\times \phi(x')^{-1} \right ) \right ] .
  \label{eq:fineq2ss2}
\end{equation}
Eq.~(\ref{eq:fineq2ss2}) is parameterized in terms of $J_{\rm eq}^0$ and the temperature of the medium. Both these quantities are 
determined by solving the initial value problem. In obtaining Eq.~(\ref{eq:fineq2ss2}), we have assumed that $J_1(0) =0$, which is permissible  as only the slope of the equilibrium solution determines the heating rate.  Eq.~(\ref{eq:fineq2ss2}) can be 
used in Eq.~(\ref{eq:ene_exchange1}) to obtain the residual heating rate in the  quasi-static state: 
\begin{equation}
    \dot Q = 4\pi c\int  dx  \left ({(h\nu_\alpha)^2 \over m_p c^2 } +{b(h\nu_{21})^2 \over 2k T_{\rm K}} \right )\sigma_0   \left[\exp(-2\eta x) \left(2C'\exp(2\eta x) \int_x^\infty dx' \psi(x') -2\gamma J_{\rm eq}^0 \right) \right] {kT_{\rm K} \over \Delta \nu_D h} .
    \label{eq:ene_exchange1ss}
\end{equation}
Here, we have assumed $T_{\rm S} = T_{\rm K}$. Our numerical results show that this rate is generally negligible, and much of the heat exchange
occurs before this quasi-static  state is reached.

The \lyal photons can both heat and cool the medium. As is known (e.g. CM04) and is discussed in detail below, the  injected photons cool, and the continuum photons heat the medium. More generally, the photon  profiles that have blueward excess  as compared to the equilibrium profile discussed above cause  heating of the medium, and the profiles that have blueward deficiency  cool the medium. The  photon profile and  the thermal state of IGM  evolve to remove the excess/deficit of photons, which drives the coupled photon-\hi\ system to equilibria 
in which both the photon distribution and the temperature of the IGM reach steady states around $x\simeq 0$.

However, it follows  from the discussion
in the foregoing that such an equilibrium might  correspond to a quasi-static state owing  to  the  expansion of the  Universe. Also, it is possible that 
the coupled system reaches a quasi-static state over different time scales depending on the spectral profile of new photons.  
For injected photons with the delta function profile, this state is reached  at $x\simeq 1$ in time $t \simeq  t_{\rm sca}/\gamma$, but it is longer by a factor of $x_{\rm max} \simeq 10^5$ for continuum photons. These diverse times  provide one  set of time scales in our analysis.

Our analysis is based on numerical solutions of Eqs.~(\ref{eq:fineq2}) and ~(\ref{eq:thermal})  along with Eqs.~(\ref{eq:ts_def}) and~(\ref{eq:ene_exchange3}).  We show that in most of the settings we study a quasi-steady state is reached. However, as anticipated from
solutions with no scattering, the  time scales for reaching this state  differ substantially.  In this state, both the photon intensity and the IGM temperature  
could only  evolve slowly  owing to the expansion of the Universe.
In the next section, we discuss these cases in detail. 

{\it A note on the choice of variables for numerical computation:} In the foregoing, we have defined independent  variables $x$ and $t'$. Both 
these variables depend on the number density of \hi\ atoms, the temperature of the medium, and the  redshift.  Even for the case of 
no scattering, the solutions (Eqs.~(\ref{eq:intdel}) and~(\ref{eq:jsolcont})) and relevant timescales appear to depend on \hi\ number density and temperature of the medium. However, in the case of no scattering, it is readily seen that number density of \hi\ atoms scales out  of Eqs.~(\ref{eq:intdel}) and~(\ref{eq:jsolcont}). A further scaling with $1/\Delta\nu_D$ converts $J(x)$ to $J(\nu)$ and there is no dependence on 
temperature in either the  photon intensity or different time scales. 

Even in the more general case, while these variables are suitable for analytic calculations, their choice 
could cause confusion if temperature changes substantially, as is the case for many scenarios we have considered here. For instance, 
as noted above,  $C$ is defined as the   number density  of 
photons  produced per unit time and solid angle in the frequency range from Lyman-$\alpha$ to Lyman-limit. For a time-independent source, $C$ is constant with no dependence on either temperature or redshift. While this variable is suitable for numerical work,  we also need to define 
other related quantities, such as $C'$, which depend on both temperature and redshift. Similarly, $x_{\rm min}$ and $x_{\rm max}$  ($x_{\rm max}$ and $x_{\rm min}$ are used in studying continuum photons; in this case, $x_{\rm max} = (\nu_\beta - \nu_\alpha)/\Delta\nu_D$ and $x_{\rm min} = -x_{\rm max}$ for our study, 
depend on temperature. To alleviate the complication of dealing with temperature dependence of independent variables, we 
compute quantities on a fixed  grid: $x_0 = x(T=T_0)$ and $t_0 = t'(T=T_0)$ where  $T_0 =10 \, \rm K$. The relevant equations 
(Eqs.~(\ref{eq:ene_exchange1}) and~(\ref{eq:fineq2})) are suitably transformed to deal with these variables. 

\section{Results} \label{sec:resl}
Before presenting our main results, we give some of the numbers relevant for our study. We use a fully neutral medium
at $z = 20$ as our fiducial setting. The scattering time scale at the line center, assuming $w = 0$,  $t_{\rm sca} = 1.5 \times 10^4 \, \rm sec$ and $1/\tau_{\rm GP} = 3.5 \times 10^{-7}$. As shown in the previous section, the steady state close to $x\simeq 0$ is reached for injected photons (with delta function injection at $x = 0$) on time scales $t_{\rm sca}/\tau_{\rm GP} \simeq 4 \times 10^{10}  \, \rm sec$ (this time is slightly longer for injected photons with  Voigt profile, RD94).  For the continuum photon,
the relaxation time scale is $x_{\rm max} t_{\rm sca}/\tau_{\rm GP} \simeq 5 \times 10^{15}  \, \rm sec$, which is nearly 160~million yr.  This time scale is comparable to the  expansion time $H^{-1} \simeq H_0^{-1} \Omega_m^{-1/2} (1+z)^{-3/2} \simeq 8\times 10^{15} \, \rm  sec$. 
This is one of the largest timescale in the problem and is comparable to  the time needed for the continuum photons
to reach quasi-static state (for detailed discussion, see Section~\ref{sec:lyah1}).

 Using $J(\nu) d\nu = J(x) dx$, it follows  that
$J(x) = J(\nu)\Delta\nu_D$. This implies that  $n_\alpha(x) = 4\pi J(x)$ is the number density of photons
in the frequency range $\nu$ and $\nu + d\nu_D$. 
At $z =20$, using the definition of $y_\alpha$ (discussion following Eq.~(\ref{eq:ts_def})) and $T_{\rm K} = T_\alpha= 10 \, \rm K$, this gives $y_\alpha \simeq 3.5 \times 10^2$  for 
$n_\alpha(0) \simeq 10^{-8} \, \rm cm^{-3}$. From Eq.~(\ref{eq:ts_def})
it follows that the spin temperature begins to get coupled to Lyman-$\alpha$ photons ($T_{\rm S} \simeq T_\alpha=T_{\rm K}$) when  $y_\alpha T_\alpha \gtrsim  T_{\rm CMB}$. For an unheated IGM at $z\simeq 20$, the onset of this condition requires $n_\alpha(0) \gtrsim 10^{-10} \, \rm cm^{-3}$, which is orders of magnitude smaller than the baryon density. 
Using Eqs.~(\ref{eq:thermal}) and~(\ref{eq:ene_exch9}), one can estimate the approximate change in the  matter temperature owing to Lyman-$\alpha$ heating/cooling:
$T_{\rm K} \simeq \dot Q/(3  H k)$.  It can be shown that this  heating/cooling is negligible  ($\lesssim 1 \, \rm K$) for the photon intensity needed to couple the spin temperature to kinetic temperature (see CM04 for details). 

In our study, we consider values of $C$ in a wide range that   gives $n_\alpha(0) \gg 10^{-10} \, \rm cm^{-3}$. It is justified based on the expected behavior of the thermal and ionization history of IGM.  The number density of baryons at $z \simeq 20$ is $n_b \simeq 10^{-4} \rm cm^{-3}$.  To ionize the neutral hydrogen  and keep it ionized,  we need around an order of magnitude larger number density  of  ionizing photons (e.g. \cite{Barkana:2000fd} and references therein). Planck
results show that the Universe ionized at $z \simeq 7.5$ so the number of ionizing photons at $z \simeq 20$ is expected to be far smaller. 
However, the number density of Lyman-$\alpha$ photons could be orders of magnitude larger than hydrogen-ionizing photons  as these photons escape from star-forming regions far  more easily\footnote{The cross section of ionization for \hi\ at threshold is $\simeq 6.3 \times 10^{-18} \, \rm cm^{2}$ while it can be shown that for the  Voigt profile the cross section for Lyman-$\alpha$ scattering falls  below $10^{-20} \, \rm cm^{2}$ for  $x \gtrsim 100$ for parameters of interest (e.g. \cite{RS18,RS19} and references therein).}. 
Therefore, many  models  corresponding to different rates of  production of 
Lyman-$\alpha$ photons at high redshifts  have been considered in the literature (e.g. CM04 and references therein). At  $z\simeq 20$, the density of these photons is essentially unconstrained  and could far exceed  the baryon density without violating Planck constraints. 

\subsection{Injected Photons}

In this section, we consider a simple model with only injected \lyal photons present in the IGM. While we expect a larger fraction of injected photons close to the sources \citep{2006MNRAS.367.1057P}, the model discussed in this section does not represent any realistic astrophysical scenario. 
We assume that the  injected photons have Voigt profile.

\begin{figure*}
  \includegraphics[width=0.67\linewidth]{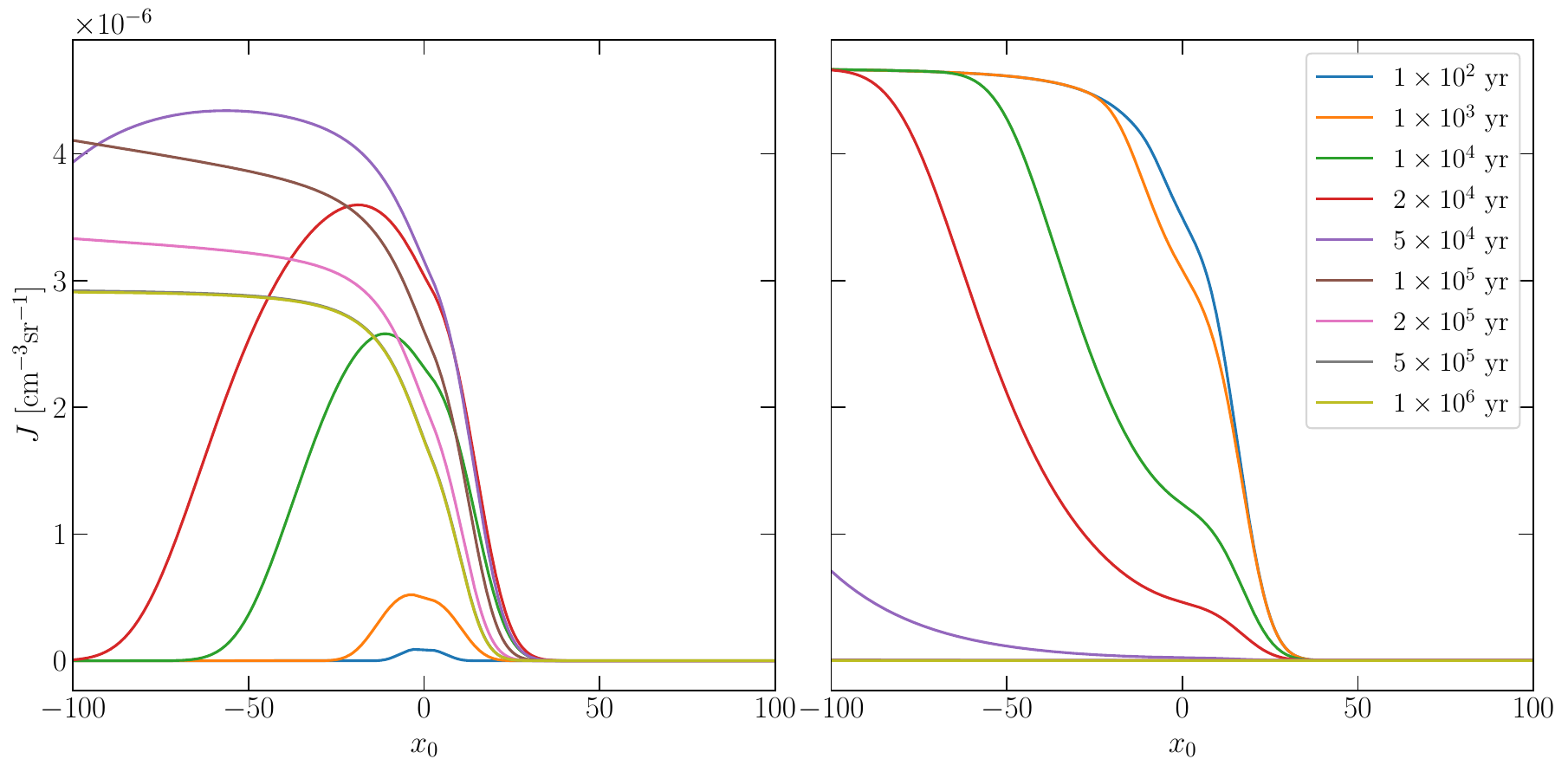}
  \includegraphics[width=0.32\linewidth]{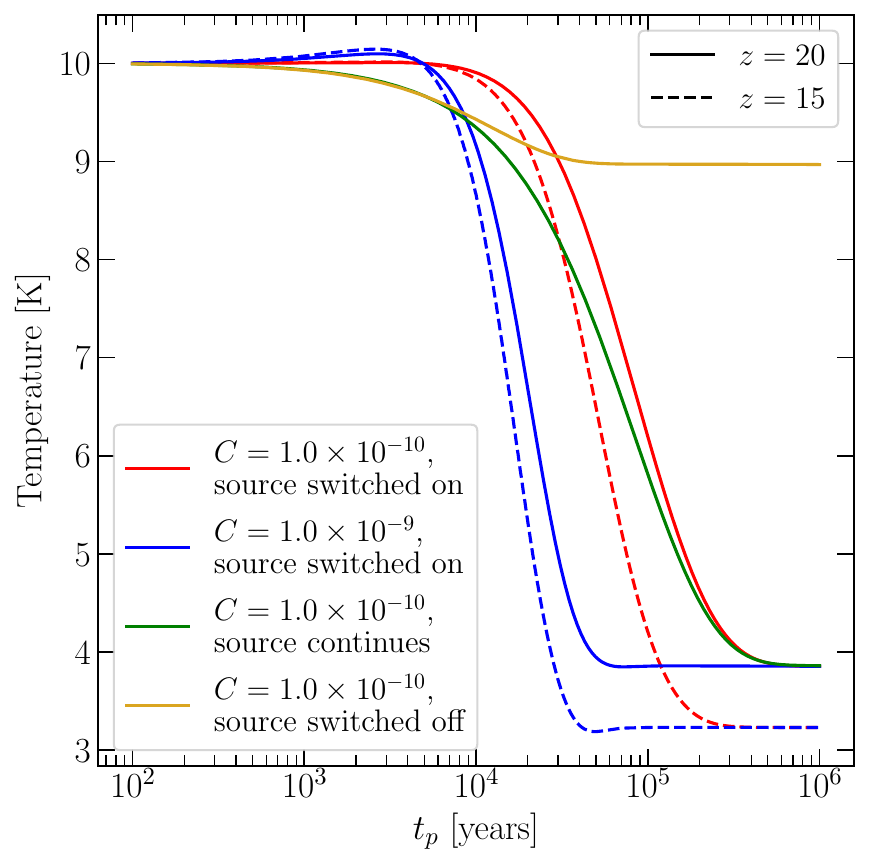}
  \caption{In the left panel, we show the evolution of intensity profile $J(x,t)$ for  photons  injected at $x = 0$ with Voigt profile. The x-axis is in units $x_0 = x(T=10 K)$. The left and the middle panel share the same y-axis (ticks on left panel) and the color scheme for time evolution of profiles (inset of middle panel). The source switches on  at redshift $z=20$ with  an initial temperature $T = 10~{\rm K}$ and source intensity  $C = 10^{-10} \, \rm cm^{-3} \, sec^{-1} \, sr^{-1}$.  In the middle panel, we start with quasi-static profile  (with the same redshift and initial temperature as the left panel) and follow the intensity profile   after switching off the source at $t=0$. In the right, panel we show the evolution of temperature for the following cases: (a) the curves `source switched on' are for sources switching on at $t= 0$ ($J(x)$ evolution is shown in the left panel). These plots correspond to two different values of $C$ (red and blue curves) and redshifts $z = 20$ (solid curves) and $z = 15$ (dashed curves),  (b) for the  case `source continues' the initial profile at $t=0$ is assumed to be the quasi-static solution at $T = 10 \, \rm K$, and it  evolves with thermal feedback, with $C = 10^{-10} \, \rm cm^{-3} \, sec^{-1} \, sr^{-1}$ and $z=20$ ($J(x)$ evolution for this case is not shown in the plot), and (c) for the case `source switched off', the initial profile is  again  the quasi-static solution with $T=10 \, \rm K$, but  the source is switched off at $t=0$ and the profile is allowed to evolve ($J(x)$ evolution is shown in the middle panel).}
  \label{fig:inj3}
\end{figure*}

In Figure~\ref{fig:inj3}, we show the evolution of the \lyal intensity $J(x,t)$ for injected photons along with the temperature evolution of the IGM (simultaneous numerical solutions of Eqs.~(\ref{eq:fineq2}), (\ref{eq:thermal}), (\ref{eq:ene_exchange3}) and~(\ref{eq:ts_def})). In the left panel of the figure, we show the evolution of the intensity profile for the case in which the source is switched 
on at $z = 20$. In our study, we choose the initial temperature $T = 10  \, \rm K$.  Our results are independent of the choice of  this temperature. At initial times, $J(x,t)$ is symmetric around $x = 0$. With time, these photons redshift, causing an
asymmetry around the resonance frequency. After nearly $t \simeq 10^6 \, \rm yr$, the profile reaches a quasi-static shape. To appreciate this time scale, we first consider the case without scattering.  Without scattering,  at $x \simeq 0$, the quasi-static profile corresponds to the balance between injection of photons and their redshifting out of the resonance (for details,
see e.g RD94, CM04; Section~\ref{sec:lyah1}). This analysis also yields  the approximate time scale at which the quasi-static profile is reached.  The scattering alters the profile close to $x\simeq 0$. We note that in this case the relaxation  time is generally  greater than $t_{\rm sca}/\gamma$ (RD94), the relaxation time-scale for delta function injection without scattering.   This is partly owing to the strong thermal feedback, as the temperature changes by more than a factor of two (right panel) and partly because the relaxation time for Voigt profile is larger than for the delta function profile (RD94).  

In the middle panels, we start with the quasi-steady-state profile  and 
study its evolution after the source is switched off. Both the left and middle panel are plotted for the same photon injection rate $C$. The intensity in  the left panel relaxes to a different value as  compared to   the middle panel, which is the expected amplitude for negligible thermal feedback. However,   the shapes of the quasi-static profiles are the same in both of the cases\footnote{The shape of the steady-state profile  in all cases can also be compared with the case without scattering discussed in the previous section (Eqs.~(\ref{eq:intdel}) and~(\ref{eq:jsolcont})).
The scattering  mainly impacts $|x|\simeq 0$ and therefore, these expressions provide the correct shape of intensity profile  for $|x| \gg 0 $.}.  This difference   demonstrates  the  impact of strong thermal feedback and  underlines an important departure  between our time-dependent analysis and the assumption of a steady-state
profile, which  does not take into account the  change in intensity  profile owing to thermal feedback (e.g. CM04, CS06). In the more general treatment we undertake in this paper,  the time-dependence 
of the photon spectrum and thermal evolution are coupled. Our work modifies and improves on earlier studies by taking this 
significant effect into account. 

In the right panel, we show the temperature evolution for  three cases: (a) In the first case, the source is switched on at $t=0$ ($J(x)$ evolution for this case is show in the left panel). (b) In the second case,  we start with a quasi-steady-state profile  at $t=0$ and let it evolve with temperature feedback ($J(x)$ evolution is not shown for this case). (c) In the third case, the source is assumed to be in quasi-static state and then switched off ($J(x)$ evolution is shown in the middle panel). The process of cooling is delayed for the first case  as compared to the third case  roughly for the period in which photon density builds in the IGM and the steady-state profile is reached. In the case where the source is switched off at $t = 0$, the temperature closely follows the steady state solution
for a similar period before the profile shifts out of the resonance, and the cooling process is  switched off. The right panel also explores
the impact of different values of photon injection rates $C$ and energy injection at different redshifts. We note that the temperature 
reaches a steady-state as the intensity profile relaxes to a quasi-static state. The steady-state temperature ,  $T \simeq 4 \, \rm K$,  is independent of $C$, at $z =20$.  We also show that, for the same $C$, the system relaxes to the same  final temperature  irrespective of whether  we switch on the source  with a Voigt profile or assume an initial  quasi-static profile at $t = 0$.  The temperature reaches a slightly smaller value 
at $z\simeq 15$. We noted above that  the hyperfine  level mixing  energy exchange (Eq.~(\ref{eq:ene_exchange})) is normally negligible.
Close to the minimum steady-state temperature, it has a  small but discernible impact (not shown in the figure); ignoring the  hyperfine energy exchange  leads to lower temperatures than shown in the figure.

The cases shown in Figure~\ref{fig:inj3} also allow us to  anticipate the outcome for a more complicated time-dependence of  photon injection. For instance, if the source is on long enough for the photon intensity and thermal state to reach quasi-steady state and is then switched off, the cooling effectively stops  in $t \lesssim 5\times 10^4 \, \rm yr$. More generally, without 
strong thermal feedback, the quasi-steady state is reached on a  time-scale $\simeq 10^5 \, \rm yr$. Even after the quasi-steady state
is reached, it is possible for the thermal state and the intensity to evolve slowly  owing to the  expansion of the Universe  or nonzero $\gamma$  (Eq.~(\ref{eq:ene_exchange1ss})). However, this possible change is too slow  to be  discernible in Figure~\ref{fig:inj3}. 

We next consider  continuum photon injection and a mix of injected and continuum photon injection and show that quasi-steady states are also reached in these cases.

\subsection{Continuum Photons}

Far away from the sources, or in presence of strong Lyman-Werner feedback \citep{1967ApJ...149L..29S, 2001MNRAS.321..385G}, we expect a scenario with only continuum \lyal photons in the IGM.
We assume the source of photons to have flat spectrum in the range $x_{\rm min} < x < x_{\rm max}$ with $x_{\rm max} \gg 0$ and $x_{\rm min} \ll 0$ (see Section~\ref{sec:lyah1} for details). For stellar sources,
there is no constraint on $x_{\rm min}$, while $x_{\rm max}= (\nu_{\beta} -\nu_{\alpha})/\Delta\nu_D \simeq 10^5$ for one branch of Lyman series (for $T_{\rm K} = 10 \, \rm K$). For the purposes of computation,
we assume $x_{\rm min} = -x_{\rm max}$. However, as argued   previously, the choice of $x_{\rm min}$ has no bearing on 
our results. 

For no scattering, the three branches  of the time-dependence of  photon intensity are  given by  Eq.~(\ref{eq:jsolcont}).  We briefly discuss the expected solutions for this case. The photons are produced  
in the frequency range $x_{\rm min} < x < x_{\rm max}$. Owing to the expansion of the Universe, after a time $t'$, these photons spread to a range $x_{\rm min} -\gamma t' < x < x_{\rm max}$. The spectral evolution 
corresponds to three different branches. In the central branch, the intensity increases linearly with $t'$
and the spectral shape is flat. We refer to this profile as the `flat profile'  for the rest of the paper. This solution has been studied in the literature  for 
continuum photon injection. However, as we show, this branch of the solution is not in a steady state.  In the frequency range $x_{\rm min} -\gamma t' < x < x_{\rm min}$, the profile is tilted and evolves with time
linearly.

Only one of the branches of Eq.~(\ref{eq:jsolcont})  corresponds to the quasi-steady state. The steady state is reached close to $x \simeq x_{\rm max}$ at early
times, and as the time progresses, the quasi-steady state reaches smaller frequency. At any time, the frequency range $x_{\rm max} - \gamma t' < x < x_{\rm max}$ is in a steady state.  In this work, we are interested in solutions close to $x \simeq 0$. The steady state reaches $x \simeq 0$  at  time $t' \simeq x_{\rm max}/\gamma$.

\begin{figure*}
  \includegraphics[width=0.67\linewidth]{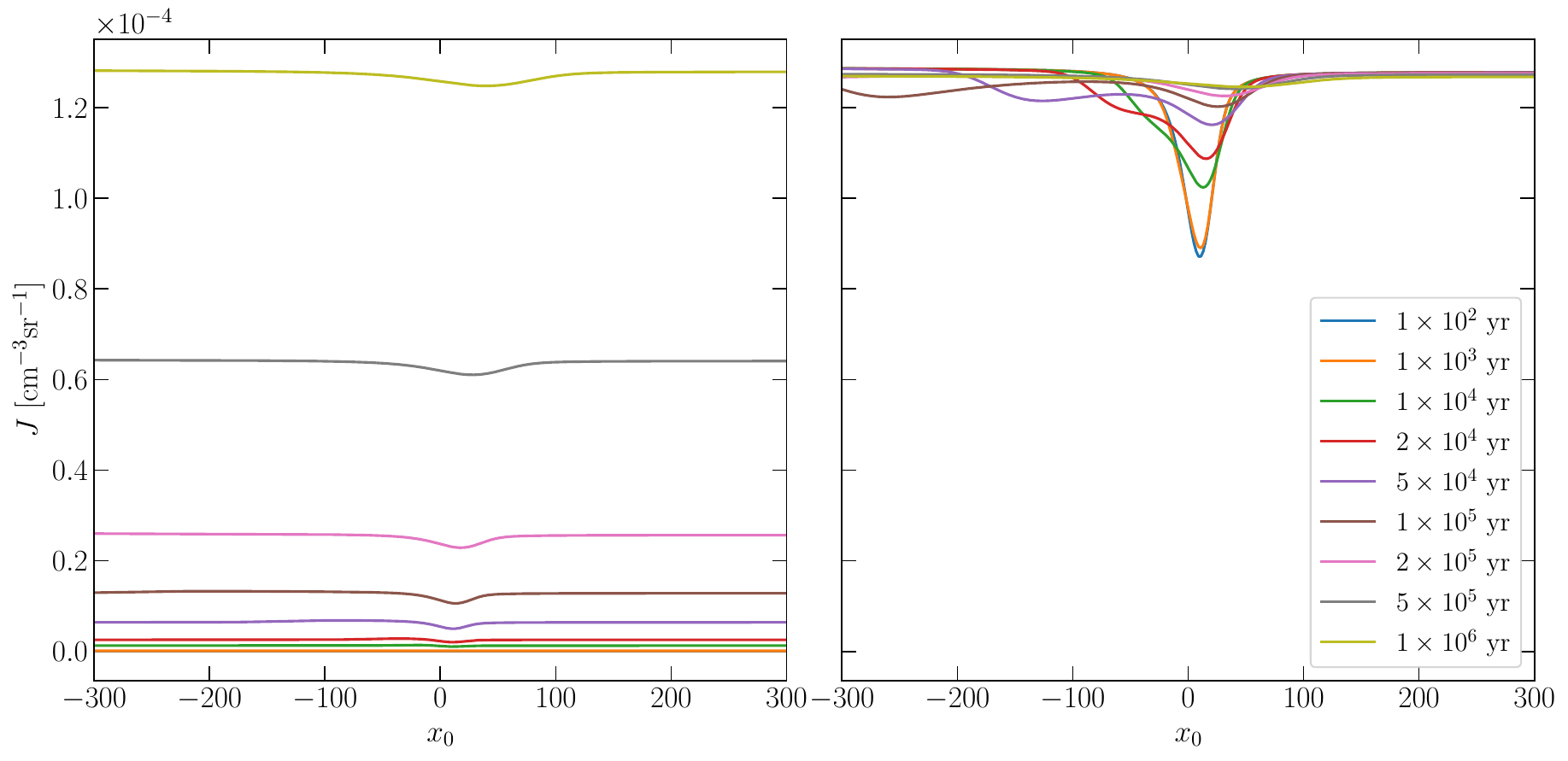}
  \includegraphics[width=0.32\linewidth]{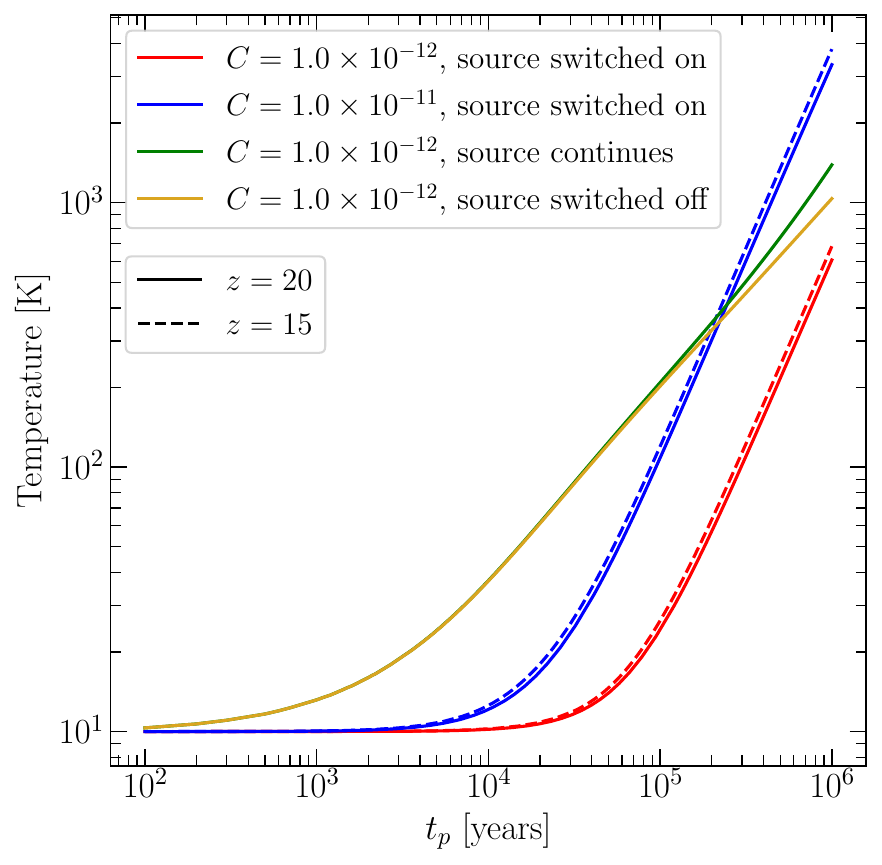}  \caption{In the left and middle panels, we show the evolution of  \lyal continuum photons intensity $J(x,t)$ at redshift $z=20$ and initial temperature $T = 10~{\rm K}$ when the source switches on (left panel) and switches off (middle panel). The energy injection rate  $C = 10^{-12} \, \rm cm^{-3} \, sec^{-1} \, sr^{-1}$.   In the right panel, we show the evolution of temperature for the same cases as in Figure~\ref{fig:inj3}.}
  \label{fig:cont3}
\end{figure*}

In Figure~\ref{fig:cont3} we show
the evolution of photon intensity  and the temperature based on simultaneous solutions of  Eqs.~(\ref{eq:fineq2}), (\ref{eq:thermal}), (\ref{eq:ene_exchange3}) and~(\ref{eq:ts_def}). In the left panel of the figure, we show the evolution of intensity 
profile for the case in which the source is switched on at $t = 0$. All the profiles shown in the left panel  correspond to the early evolution of the photon intensity, $t \ll t_{\rm sca} x_{\rm max}/\gamma$.  
The main impact of scattering is to alter the intensity pattern close to $x\simeq 0$.   At early times, $x \simeq 0$ falls on the flat asymptote in Eq.~(\ref{eq:jsolcont}) and the intensity increases linearly in this regime. This corresponds to the  case studied in the literature (e.g. CM04, CS06). However, this branch of the solution does not
correspond to the quasi-steady state solution.  In the middle panel, the evolution of intensity profiles is shown when the source 
is switched off at $t = 0$; the  initial profile of the source is assumed to be the flat profile studied in the literature (e.g. CM04). 

The temperature
evolution  for many cases  is studied in the right panel of the figure; it can be summarized as follows:
\begin{itemize}
\item[(a)] The heating gets delayed if the source is switched on at $t = 0$ as compared to the `flat  profile' injection of 
energy (starting profile of the middle panel) but rises more sharply at later times. The temperature rise tapers off for the case
in which the source is switched off.  We discuss this case in more detail later. 
\item[(b)] The most striking difference between the  right panels of Figure~\ref{fig:inj3} and~\ref{fig:cont3} is that the  temperature 
does not reach an equilibrium for continuum photon profiles shown in the  left and middle panels of Figure~\ref{fig:cont3}. It is expected
from our argument that flat profiles don't  correspond to a quasi-static  intensity profile. 
\item[(c)] Figure~\ref{fig:cont3} also explores the impact of higher photon injection rate. For 
a larger $C$, the temperature rises more sharply. We also note that the redshift dependence  of the temperature is mild. 
\end{itemize}

\begin{figure*}
  \includegraphics[width=0.67\linewidth]{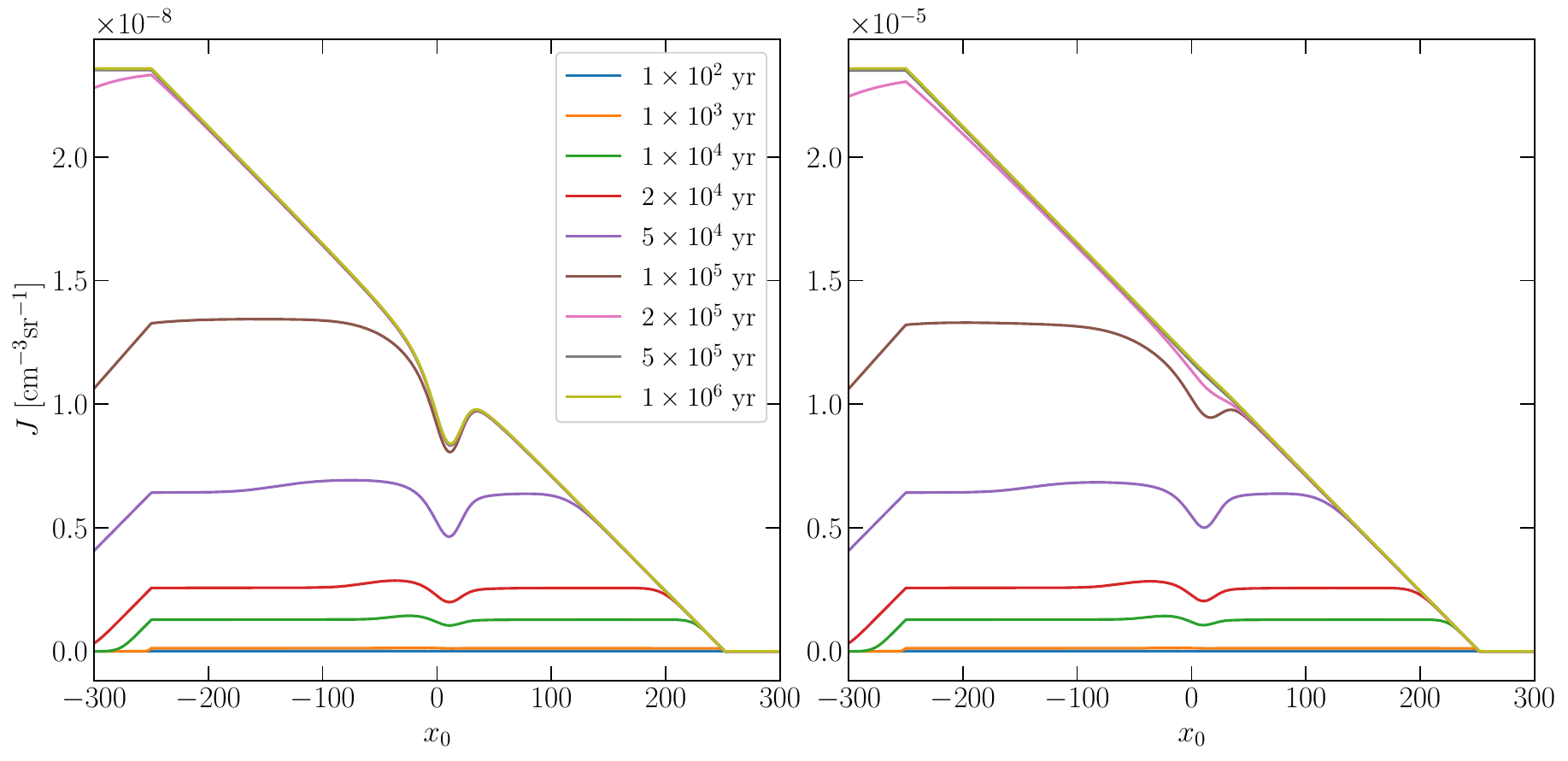} 
  \includegraphics[width=0.32\linewidth]{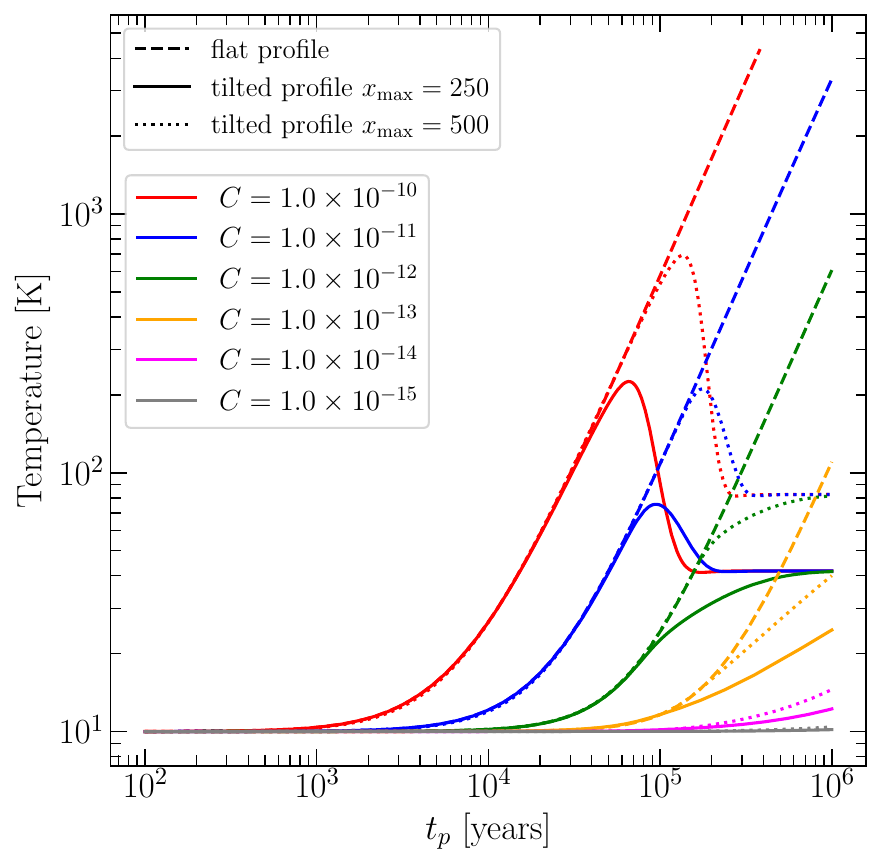}   
  \caption{This figure shows all the three branches the intensity profile $J(x,t)$ for continuum energy injection, including the approach to quasi-static intensity profile (Eq.~(\ref{eq:jsolcont})). In all the panels,  $x_{\rm max} = 250$,  the initial temperature $T = 10~{\rm K}$, and $z = 20$. The left and the middle panels correspond, respectively, to $C = 10^{-15}\, \rm cm^{-3} \, sec^{-1} \, sr^{-1} $  and $C = 10^{-12} \, \rm cm^{-3} \, sec^{-1} \, sr^{-1}$. In the right panel, we show  the   temperature evolution   for a range of photon injection rates,  $C$ (solid curves), which includes the two cases for  which the intensity profiles are shown in the left and the middle panel. We also show the temperature evolution for the flat profiles displayed in Figure~\ref{fig:cont3} (dashed curves) and the tilted profile with $x_{\rm max} = 500$ (dotted curves) for comparison. $J(x)$ evolution for $x_{\rm max} = 500$ case is not shown here. We see that the final equilibrium temperature is a function of $x_{\rm max}$, with larger $x_{\rm max}$ resulting in higher  final temperature.}
  \label{fig:cont_numax}
\end{figure*}

We next discuss the approach to the quasi-static  photon profile and its implications. As noted above,  $x_{\rm max} \simeq 10^5$ in the realistic case. However,  it takes close to Hubble time to reach the quasi-steady state for such a large value of $x_{\rm max}$ and it 
is computationally expensive to integrate for such a long period. To illustrate the convergence to a quasi-steady state solution, we choose $x_{\rm max} = 250$ in Figure~\ref{fig:cont_numax}.  

The intensity profiles  in Figure~\ref{fig:cont_numax}
should be compared to the no-scattering solutions (Eq.~(\ref{eq:jsolcont})). As expected, the main impact of the scattering
is to alter the intensity around $x \simeq 0$ and for $|x| \gg 0$ the profile approaches the no-scattering solutions. All three
branches of solutions given in Eq.~(\ref{eq:jsolcont}) are seen in this figure. 
In the left/middle panel of the figure, the approach to the quasi-steady state branch of the solution is shown for cases when
the thermal feedback is negligible/appreciable. For both of the cases, we see convergence to a near  tilted profile in accordance with 
the expectation  based on Eq.~(\ref{eq:jsolcont}), even though the final profile close to $x \simeq 0$ differs. 

The temperature evolution based on photon intensity profiles  for  $x_{\rm max} = 250$  is displayed in the right panel of Figure~\ref{fig:cont_numax} for 
different values of $C$. In contrast to the flat profile case, we find  an equilibrium temperature is reached in all cases in Figure~\ref{fig:cont_numax}. We follow  the temperature evolution for $t = 10^6 \, \rm yr$ in the figure. During this period, the temperature has already reached a quasi-static state for $C \gtrsim 10^{-12} \, \rm cm^{-3} \, sec^{-1} \, sr^{-1}$ but it is still evolving for smaller rates of injection. This behavior is partly in line with our expectation based on the no-scattering solution  that the temperature reaches a constant when the photon  intensity  relaxes to a tilted
profile. It is also seen that the final temperature is independent of the photon injection rate $C$, as we earlier found for 
the injected photon case (right panel of Figure~\ref{fig:inj3}).  The left and middle panels allow us to compare  the approach to equilibrium  as seen in the right panel. In the left panel, $C = 10^{-15} \, \rm cm^{-3} \, sec^{-1} \, sr^{-1}$, which is too small to allow an approach to thermal equilibrium even after the profile becomes tilted.
The profile close to $x \simeq 0$ for such a small value of $C$ should be contrasted with the one in the middle panel for which $C$ is three orders of magnitude 
larger. In this case, the profile close to $x\simeq 0$ approaches the equilibrium profile $\propto \exp(-2\eta x)$, which drives the temperature to its equilibrium value. 

In the right panel of Figure~\ref{fig:cont_numax}, we also display the thermal evolution for $x_{\rm max} = 500$. In this case, the final temperature is higher.  This is expected as a larger $x_{\rm max}$ means the photon profile remains flat for longer before the onset of the tilted profile. The flat profile cases are also shown in the Figure for comparison. Here, the IGM keeps heating so long as the profile remains flat. This results in a larger equilibrium temperature for higher $x_{\rm max}$.

Based on our  analysis for injected and continuum photons, we can identify key determinants of the time-dependence of the coupled system. The temperature evolution depends on a number of parameters: redshift, temperature, the photon profile shape,  and the number of photons in the IGM. We have  seen that  in both the cases the temperature starts rising or falling slowly when the source is switched on.  Once a suitable profile shape has been reached and the intensity has built to large enough values, the temperature starts evolving faster owing to a rapid exchange of energy between photons and neutral hydrogen (Eq.~(\ref{eq:ene_exchange1})). This  also causes  the photon profile 
to relax to its equilibrium profile. The net impact of these two processes is to drive the  rate of energy exchange $\dot Q$ to a very small value and the  temperature to  a quasi-static state (Eqs.~(\ref{eq:thermal}) and~(\ref{eq:ene_exchange1})).  This occurs for all cases we study except  the flat continuum photons profile, which is not an equilibrium profile. The time scale over which   the  temperature reaches quasi-static state  depends on the number density of photons, which  in turn depends on the rate at which the new photons are injected. 

Both Figures~(\ref{fig:inj3}) and~(\ref{fig:cont_numax}) show that, for a larger value of $C$, the equilibrium temperature is reached faster as the higher  number density of photons increases the rate of energy exchange with the medium (Eqs.~(\ref{eq:thermal}) and~(\ref{eq:ene_exchange1})). For a very large value of $C$, the temperature   overshoots the equilibrium point  before eventually settling at the quasi-static value.  This is   owing to the adjustment of  time scales  at  which  photons profiles  heat/cool the medium and the back reaction of the thermal state on these profiles. The expansion rate does not play an important role in these cases. 

For much smaller values of $C$, as seen in the right panel Figure~(\ref{fig:cont_numax}), the temperature continues to evolve  even though the intensity at large $x$ has reached  a quasi-static state.  In such cases, the photon intensity in the medium is not large enough for a rapid exchange of energy. A  comparison of  intensity profiles at $x\simeq 0$ between the left and middle panels of Figure~(\ref{fig:cont_numax}) underlines the difference between large and small $C$. While the profile for large $C$ has reached close to  the equilibrium form, the profile for small $C$ is yet to reach that form. For very small $C$, the time scale to reach the equilibrium  could be comparable to the expansion time scale of the Universe and further  evolution in these cases is determined 
by a balance between the heating and the expansion of the Universe. 

In our study, we use $x_{\rm max} = 250$ and~$500$ to illustrate the approach to equilibrium of the coupled photon-gas system. However, $x_{\rm max} = 10^5$ in the realistic case. In the next  section, we discuss how the transition from the flat to tilted profile could impact the final temperature. We  expect this effect to be further cooling, and  the extent of cooling  is expected to be small, but  it could depend on the duration of 
the tilted profile. This suggests that we can approximate  the  long-term evolution of the system   by scaling  the time scales on the x-axis by a  factor $10^5/250$. 
We point out one possible implication for such a large value of $x_{\rm max}$. Figure~\ref{fig:cont3} shows that, for a continuum flat profile case, the temperature  keeps rising. This increase continues  for $t\simeq t_{\rm scat} x_{\rm max}/\gamma$, the time during which the profile remains flat for $x \simeq 0$.
If the temperature continues to rise at the rate shown in the right panel of Figure~\ref{fig:cont_numax}, it could reach arbitrary high values, e.g.,  the ionization temperature of \hi. However, as already discussed in the literature, the production of continuum photons without injected photons does not correspond to the  physical situation in the IGM. We turn to this case next.

\subsection{Continuum and Injected Photons}

\begin{figure*}
  \includegraphics[width=0.67\linewidth]{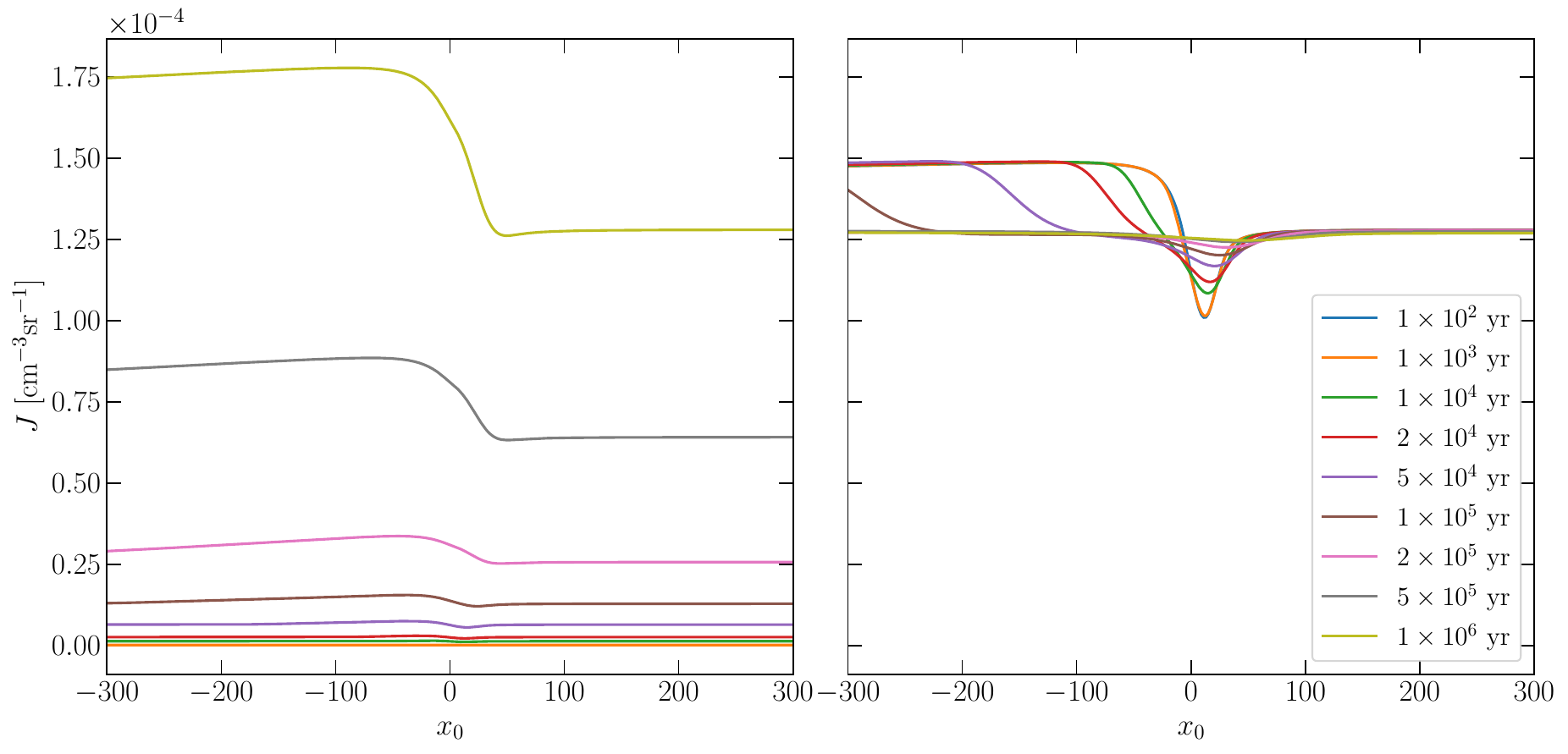}
  \includegraphics[width=0.32\linewidth]{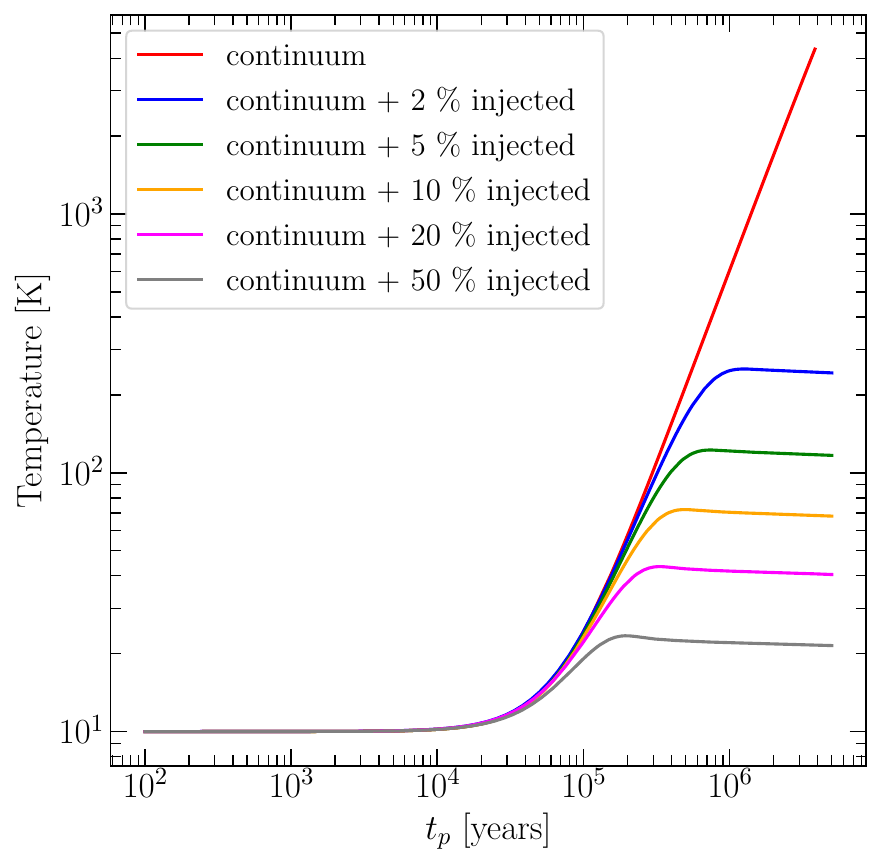}
  \caption{In the left and middle panels, we show the evolution of  photon intensity profile for  the  continuum + 20\% injected photons case  at redshift $z=20$ with an  initial temperature $T = 10~{\rm K}$ and $C = 10^{-12} \, \rm cm^{-3} \, sec^{-1} \, sr^{-1}$. In the left panel,  the source switches on at $t=0$; while in the middle panel, the initial profile is assumed to have  quasi-static form, and it  switches off at $t=0$. The middle panel shows that,  when the source is switched off, the injected feature redshifts out of the resonance  much faster than the continuum feature. Therefore, this scenario heats up the gas (see the middle and right panel of Figure~\ref{fig:cont3} for comparison). In the right panel, we show the evolution of temperature  for different injected photon fractions and compare it with the continuum photon only case. We see here that the temperature of the gas reaches different equilibrium values for different injected photons fraction for the same value of $C$. }
  \label{fig:cont+inj0.2}
\end{figure*}

In a more realistic case, both continuum and injected photons are expected in the IGM. Injected photons arise from the absorption of 
Lyman-series photons to  energy levels  $n > 3$ and these photons could  contribute nearly 10-20\% of the continuum photons on average. The  ratio of injected to continuum photons depends on the stellar spectrum and approaches 0.2 for a hot source (e.g. \cite{chuzhoy2007heating}; see also \cite{chuzhoy2006ultraviolet,hirata2006wouthuysen,furlanetto2006scattering}). However, the exact ratio depends on the distance from the source, with the larger fraction of injected photons close to the source.

In Figure~\ref{fig:cont+inj0.2}, we show the photon  intensity
and thermal evolution for  the case in which both continuum and  injected photons are present.  The excess of photons on the redward side  of intensity profiles is an expected feature of the intensity profile in this case and follows  from adding the intensity profiles of injected and continuum photons (Figures~\ref{fig:inj3} and~\ref{fig:cont3}).  The left and the middle panels show the intensity profiles
for the cases studied earlier (switching on and switching off sources and starting with near quasi-static  profile (Figures~\ref{fig:inj3} and~\ref{fig:cont3})). In the right panel  of the Figure~\ref{fig:cont+inj0.2}, the thermal evolution is shown  for many cases when the fraction of   injected photons is varied.

\begin{figure*}
  \includegraphics[width=1.0\linewidth]{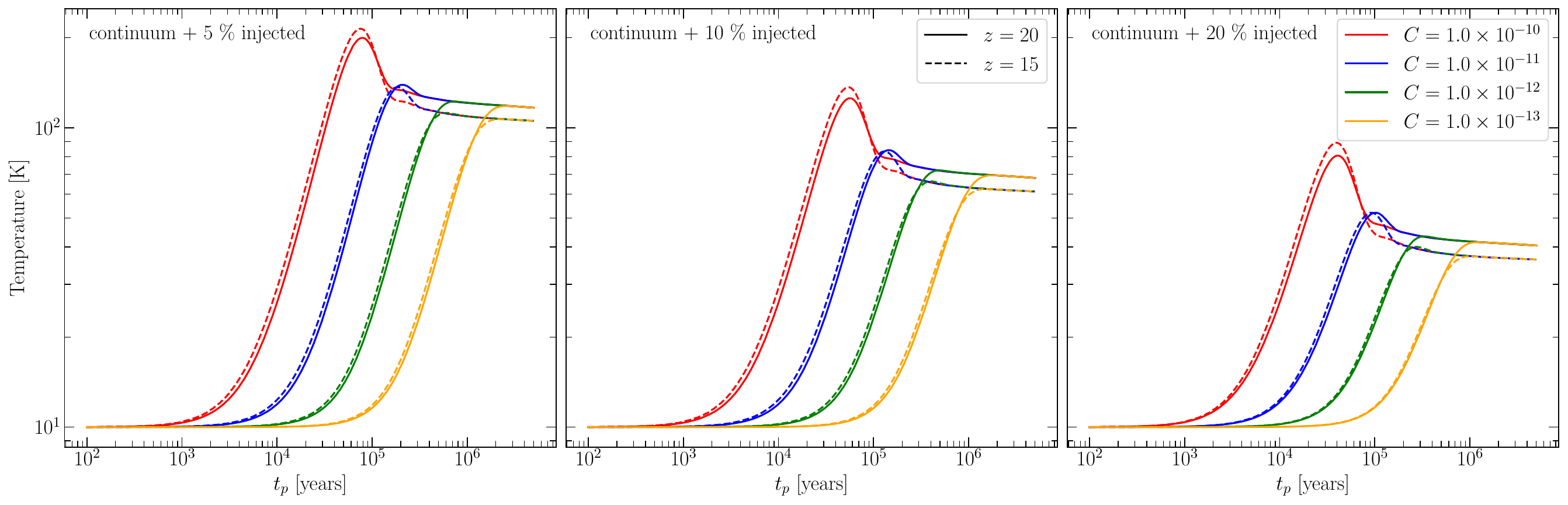}
  \caption{The temperature evolution is displayed for 5\% (left), 10\% (middle), and 20\% (right) fraction of  injected photons for initial temperature $T_{\rm K} = 10~$K and two redshifts. The solid curves are for $z = 20$ and dashed curves are for $z=15$. For higher fraction of injected photons, the equilibrium temperature  is lower. This final equilibrium temperature does not depend on the rate of energy injection $C$, but only on the ratio of the continuum to injected photons. It also does not depend on the initial temperature (we have not shown it here). The final temperature has a mild dependence on redshift, with lower  temperature reached at smaller redshifts.} 
  \label{fig:cont+inj_temp}
\end{figure*}
As the injected profile cools and the flat, continuum profile heats, a 
mix of continuum and injected photons results in lower heating rates as compared to the flat, continuum case. As we have seen in the foregoing, the time scales over 
which the injected and continuum photons reach steady-state profiles are very different, so we expect the combined profile 
to evolve over the longer of the two time scales. From the left panel of  Figure~\ref{fig:cont+inj0.2}, we notice that the profile
is still evolving after 1 million yr, which is the expected behavior for a flat, continuum profile.  In the middle panel of Figure~\ref{fig:cont+inj0.2}, we notice an important aspect of mixing the two profiles: after the source is switched off, the injected profile disappears as it redshifts to negative values of $x$ on  a time scale much shorter than the continuum profile. We discuss the implication of this feature in the next subsection.  
One important outcome of the
thermal evolution of IGM for a mix of injected and continuum profiles is that the temperature reaches equilibrium in a million yr for $C \gtrsim 10^{-12} \, \rm cm^{-3} \, sec^{-1} \, sr^{-1}$, with the equilibrium
temperature lower for a larger fraction of injected photons (right panel of Figure~\ref{fig:cont+inj0.2}). 
We explore the thermal evolution of the gas further 
for different rates of photon production, $C$, and  ratios of injected  to continuum photons in Figure~\ref{fig:cont+inj_temp}. 
All the cases shown in Figure~\ref{fig:cont+inj0.2} and~\ref{fig:cont+inj_temp} correspond to intensity profiles that are 
evolving (left panel of Figure~\ref{fig:cont+inj0.2}).

\begin{figure*}
    \includegraphics[width=0.67\linewidth]{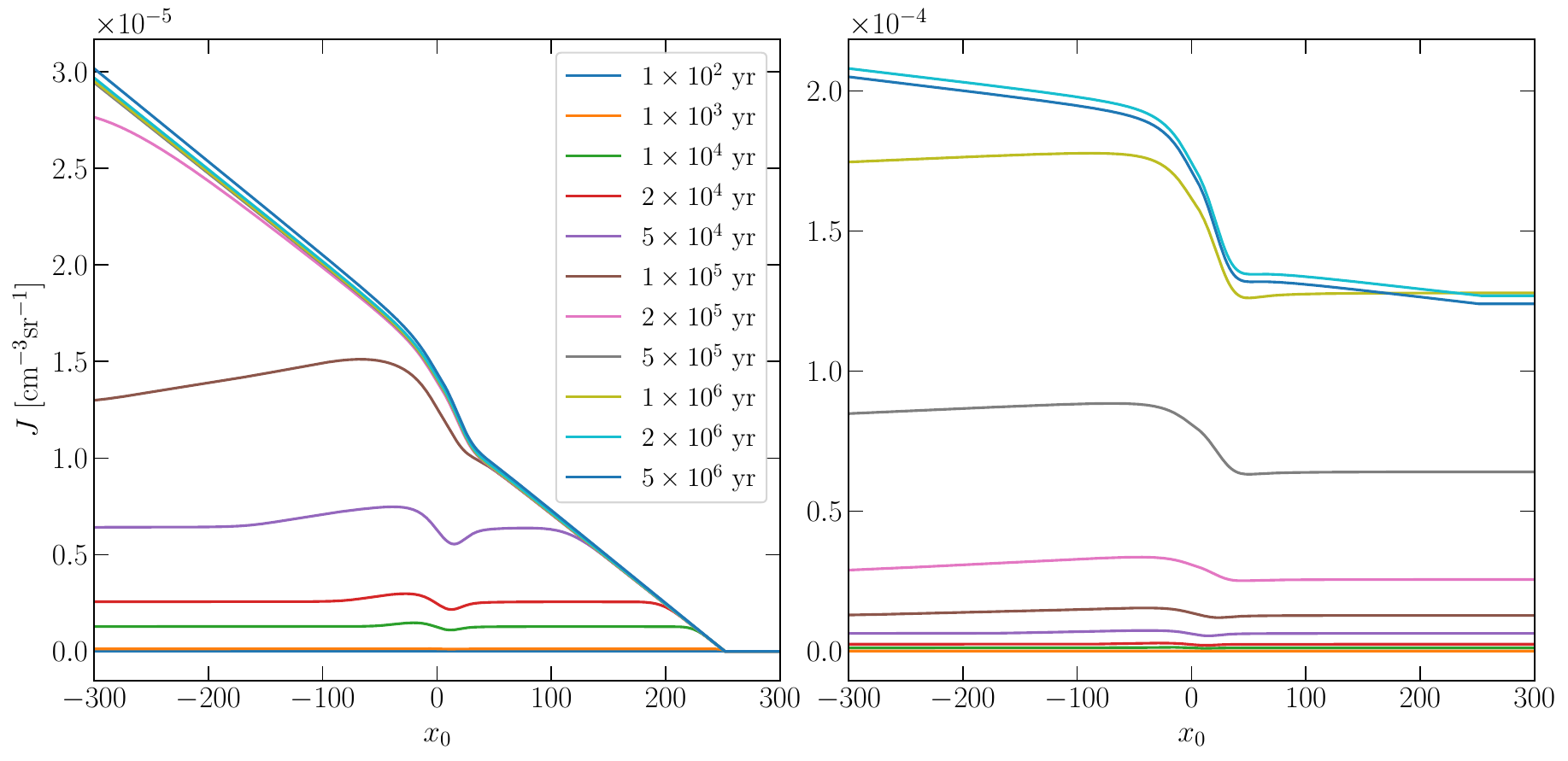}
    \includegraphics[width=0.32\linewidth]{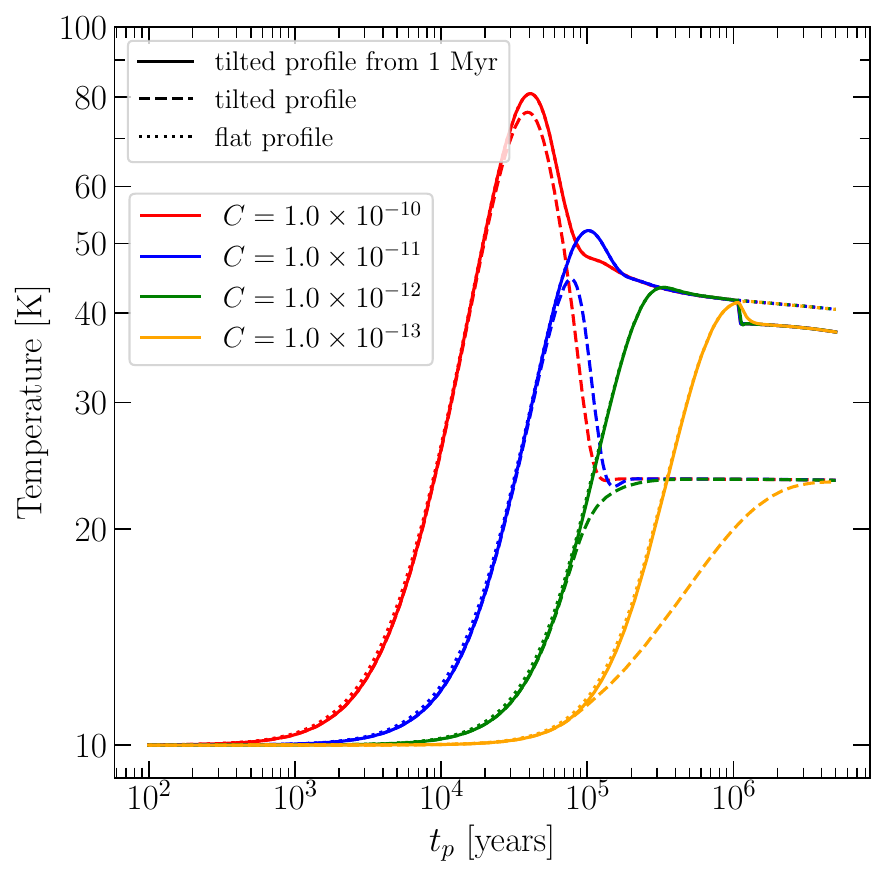}
  \caption{This figure explores the approach from the flat to the  tilted profile on the thermal evolution of the IGM. It corresponds to \lyal continuum + 20\% injected photons  at  $z=20$ and initial temperature $T = 10~{\rm K}$. As in Figure~\ref{fig:cont_numax}, we  model the approach to tilted profile by assuming $x_{\rm max} \ll 10^5$. In the left panel, we start with the tilted profile from $t=0$. In the middle panel, we start with a flat profile, and  the tilt starts at after 1~Myr. In the right panel, we show the comparison of evolution of temperature with time in these two cases, along with the flat profile case (right panel of Figure~\ref{fig:cont+inj_temp}; dotted curves). The dashed (solid) curves  correspond to thermal evolution for intensity profiles of  left (middle) panel. The temperature in all three cases  reach equilibrium  with dotted curves (which follow the solid curves for $t \lesssim 1 \, \rm Myr$) having the highest temperature and the dashed curves with  the lowest temperature. A comparison between dotted and solid curves   shows  that the  heating rate is lower for a tilted profile, which is further underlined by dashed curves for  which the tilted profile is switched on at $t=0$. }
  \label{fig:cont+inj_ssint}
\end{figure*}

In Figure~\ref{fig:cont+inj_ssint} we study the approach to equilibrium, a quasi-static state in which both temperature and  intensity profile are non-evolving. 
This figure is akin to Figure~\ref{fig:cont_numax}. As in Figure~\ref{fig:cont_numax}, we assume $x_{\rm max} \ll 10^5$. While this assumption  has no bearing on the equilibrium  states seen in Figures~\ref{fig:cont+inj0.2} and  \ref{fig:cont+inj_temp}, it impacts the long-term evolution of the system. As in the continuum case,  we assume that the time axis of the right panel can be  scaled by a factor of  a few  hundred  to capture this dynamics. Figure~\ref{fig:cont+inj_ssint} allows us to partly assess the validity of this assumption, by exploring the dependence of the final temperature on the time scale over which the flat profile makes a transition to the correct equilibrium tilted profile. In Figure~\ref{fig:cont_numax}, we explored this dependence  by comparing  two different  values  of $x_{\rm max}$ for continuum photons. We see that, in both figures, the final equilibrium temperature depends on the time scale over which the $J(x)$ profile near $x = 0$ transitions from the flat to tilted profile. The larger $x_{\rm max}$ increases this time, and it leads to larger values of the final temperature.

Figures~\ref{fig:cont+inj0.2}, \ref{fig:cont+inj_temp},  and~\ref{fig:cont+inj_ssint} display some of  the key results of this paper.  These can be summarized as follows:
\begin{itemize}
\item[(a)] Unlike the continuum case, for which the temperature continues to rise for the evolving (flat)  intensity profile (right panel of Figure~\ref{fig:cont3}), a mix of continuum and 
injected photons causes the temperature to reach equilibrium even for such profiles. 
\item[(b)] The thermal evolution of the gas depends strongly on the ratio of  injected to continuum photons, with a lower equilibrium temperature for a higher percentage of injected photons. The equilibrium temperature is lower at smaller redshifts (Figure~\ref{fig:cont+inj_temp}).
\item[(c)] The equilibrium value of temperature is  independent of the rate of photon production, $C$.  We verify this result in Figure~\ref{fig:cont+inj_temp} by varying $C$ by 4 orders of magnitude. 
\item[(d)] Both (a) and~(b) were partly anticipated by \cite{chuzhoy2006ultraviolet,chuzhoy2007heating} based on steady-state solution of intensity. Our work extends their analysis  and  allows  us to estimate important time scales. 
Figure~\ref{fig:cont+inj_temp} shows   that the steady-state temperature is reached within  a   million yr for $C \gtrsim 10^{-12} \, \rm cm^{-3} \, sec^{-1} \, sr^{-1}$.
As the equilibriation time scale for the intensity profile  is $t' \simeq \gamma x_{\rm max} \gg 10^6 \, \rm yr$, the 
thermal equilibrium is reached when $x \simeq 0$  is still on the flat, and therefore evolving,  branch  of the profile (Figures~\ref{fig:cont_numax}). 
We also note that the final equilibrium  temperature is nearly independent of the initial temperature (not shown).  
\item[(e)] From nearly a few million yr, the time scale over which the thermal  equilibrium shown in  Figures~\ref{fig:cont+inj0.2} and~\ref{fig:cont+inj_temp} is reached,  to around 0.1~Gyr after switching on the source, the temperature  is constant, but the intensity profile
evolves around $x \simeq 0$. After 0.1~Gyr, the intensity profile makes a transition from the flat spectral shape to a tilted one and reaches the steady state at $x \simeq 0$. The transition between these two periods 
on the temperature evolution  is captured in Figure~\ref{fig:cont+inj_ssint}.  A comparison between Figures~\ref{fig:cont+inj_temp} and~\ref{fig:cont+inj_ssint} shows 
that the net impact of transition from the flat to the tilted profile is further cooling, followed by the temperature reaching a 
constant value. The final state corresponds to both the temperature and the photon profile reaching quasi-static states. 
\item[(f)] Figure~\ref{fig:cont+inj_ssint} suggests that Lyman-$\alpha$ photons can heat the IGM to a maximum temperature of around 
50--100~K, depending on the fraction of injected to continuum photons (e.g. \cite{chuzhoy2006ultraviolet,chuzhoy2007heating}).  
\end{itemize}

\subsection{Long-term Evolution of the coupled \texorpdfstring{Photon-\hi\ }{Photon-HI }system}
For both the continuum and continuum+injected photon cases, we assume $x_{\rm max} \ll 10^5$ to capture the long-term evolution of the 
system.  What are the limitations of this assumption? For large enough photon production rate $C$,  in short-term ($t \lesssim 1 \, \rm Myr$), the injected profiles and the continuum+injected profiles cause the system to reach thermal equilibrium. We verify that photon 
profiles reach close to the equilibrium profile $\propto \exp(-2\eta x)$ close to $x = 0$ in these cases, which drives the energy 
exchange rate $\dot Q \simeq 0$ (Eqs.~(\ref{eq:thermal}) and~(\ref{eq:ene_exchange1})). This is also true for the tilted profile 
in the continuum case (Figure~\ref{fig:cont_numax}). During this phase, the  adiabatic cooling  owing to the expansion of the Universe (Eqs.~(\ref{eq:thermal})) does not play an important role. For smaller $C$, the thermal state is still evolving at $t \simeq 1 \, \rm Myr$.

What is the expected evolution of the thermal state of IGM  between  1~Myr and 100~Myr, the period over which the continuum photon profile continues to evolve? The injected profile reaches a quasi-static form in less than a million yr (left panel of Figure~\ref{fig:inj3}) and the residual energy exchange between photons and the \hi\ is driven by the expansion of the Universe (for details, see Section~\ref{sec:lyah1}, Eq.~(\ref{eq:ene_exchange1ss})), which, our results show, is small. The photon profile for the other cases evolves during this 
period. However, our numerical work with smaller $x_{\rm max}$ shows we do not expect any significant change in the thermal state during 
this period, in particular for large $C$ for which the thermal equilibrium is reached within a million year. For smaller $C$, we do expect
a more complex thermal evolution, especially after the expansion rate of the Universe starts playing an important role. 

Our main  inference  is based on solving an initial value problem based on a  time-independent source (constant $C$)  with the spectral shape given by injected or continuum photons. We also consider   more complicated source histories  in our analysis, e.g. starting with a quasi-static profile or switching off the source (the middle and right panels of Figures~\ref{fig:inj3}, \ref{fig:cont3}, \ref{fig:cont+inj0.2}).
We next discuss the long-term thermal evolution of a source that is switched off after 0.1--1~Mpc. The evolution of spectral profile for such a scenario is shown in the middle panels of Figures~\ref{fig:inj3}, \ref{fig:cont3}, and  \ref{fig:cont+inj0.2} for different cases.  For injected photons, the photons redshift away from $x\simeq 0$, thereby shutting off the interaction between photons and atoms, which causes the temperature to freeze (right panel of Figures~\ref{fig:inj3}). For continuum photons, the redshifting of the pattern 
brings photons into the resonance from the blueward side, which continues the heating, as seen in the right panel of Figure~\ref{fig:cont3}\footnote{The dilution of photon density owing to expansion is negligible as the relevant time scale is far shorter than the expansion rate.}. For a mix of injected and continuum photons, the injected photons redshift away from $x=0$, leaving only
the continuum photons close to the resonance frequency (middle panel of Figure~\ref{fig:cont+inj0.2}). This means the IGM keeps heating,
and it does not relax to equilibrium. In this case, the temperature would exceed the equilibrium temperature. The heating could continue
until the continuum photons redshift to the redward side of the resonance. This time scale depends on the age of the source and how fast the pattern close to $x = 0$ alters owing to photon free-streaming, etc. This 
example shows how short-lived sources could have nontrivial impact on the thermal evolution of the system. We will  consider more complex 
source histories and their  impact of  long-term  thermal evolution  of the medium in a later work. 

Most theoretical analyses related to \hi\  observables  ignore the role of Lyman-$\alpha$ heating. The main role of Lyman-$\alpha$ photons
is to couple the spin temperature of \hi\ to Lyman-$\alpha$ colour temperature through Wouthuysen-Field effect (Eq.~(\ref{eq:ts_def})) \citep{wouthuysen1952excitation, field1958excitation}; the Lyman-$\alpha$ colour temperature  is further assumed to 
equal the kinetic temperature \citep{field1959time}. As is known, and also noted above, the Lyman-$\alpha$ photon density needed to cause  this coupling results in negligible heating. 
In this work, we have considered models with larger Lyman-$\alpha$ photon  densities, which could yield substantial heating.

We consider homogeneous heating, which is a good assumption  as the mean free path of 
Lyman-$\alpha$ photons (time taken to redshift a Lyman-$\beta$ photon into a Lyman-$\alpha$ photon) exceeds a few hundred Mpc (at $z \simeq 20$ this distance (comoving) is $\simeq 300 \, \rm Mpc$, e.g. \cite{RS18} and references therein). As the  Lyman-$\alpha$ heating  could raise
the temperature of the IGM above the CMB temperature (Figure~\ref{fig:cont+inj_temp}), it could be the sole source of heating before 
X-ray heating becomes important and complement X-ray heating at late times. One important outcome of 
our study is that, for large enough photon injection rates, $C$, an equilibrium temperature is reached over
a few million yr,  independent of $C$ (Figure~\ref{fig:cont+inj_temp}). Given the ratio of injected to continuum photons (which depends on
stellar spectra), such scenarios can be incorporated in low-time resolution CD/EoR \hi\ codes by raising the 
IGM temperature to the equilibrium temperature. The equilibrium temperatures for 
different ratios of injected to continuum photons are  given in Figure~\ref{fig:cont+inj_temp}. In Figure~\ref{fig:cont+inj_ssint} we show that the late-time evolution of IGM only lowers this temperature
marginally.

The situation is more complicated at shorter times  as the evolution of thermal profile in IGM owing to Lyman-$\alpha$ heating  depends on specific models, as shown in Figures~\ref{fig:inj3}, \ref{fig:cont3}, \ref{fig:cont+inj0.2}. This is further complicated if  the source is switched off as this could result in the resumption of heating. Therefore, it is difficult to infer generic features of IGM heating in such cases. The more general treatment of this issue  depends on the interplay of time-dependence of  $C$ and the ratio of injected to continuum photons,  along with long-term evolution of the system. We have analysed only a part of this parameter space in this paper. In a later work, we hope to explore the full range of relevant 
 parameters along with  specific models of Lyman-$\alpha$  photon injection. 

\section{Conclusions} \label{sec:conc}
In this paper, we study the thermal impact of Lyman-$\alpha$ photons during the era of cosmic dawn.  Our work extends earlier 
analyses by studying the  fully  time-dependent, coupled system of Lyman-$\alpha$  photons and the thermal state of the IGM. This allows us to both verify earlier results based
on quasi-steady state solutions  and determine the important time-scales in the problem.

We encounter a diversity of time scales in our study. In particular,  we identify  four important time scales that have a bearing on our results: 
\begin{itemize}
\item[(a)] \textit{The time taken for photons to redshift from Lyman-$\beta$ to \lyals.} This time scale is comparable to the expansion time scale of the Universe and it affects only the continuum spectral profile. It yields the time scale over which  the  evolving flat profile
reaches the time-independent tilted profile. This time scale, on the order of  $t_{\rm scat}/\gamma x_{\rm max}$,  is a few hundred million yr and depends only on  the redshift. 

\item[(b)] \textit{The time taken for the profile to reach its quasi-static equilibrium value within one Doppler width.} This time scale is $\simeq t_{\rm sca}/\gamma$  and depends on the temperature of the medium and the redshift.  It also provides the approximate time over which the injected profile reaches quasi-static state without significant thermal feedback.  

\item[(c)] \textit{The time taken to reach the equilibrium temperature.} This time scale  depends on redshift, temperature, and the photon production rate ($C$). Larger production  rate results in quicker   photon number density build up  in the medium, which causes faster  exchange  of energy between photons and neutral hydrogen.  This time scale could  differ significantly  from the time taken to reach the quasi-static   photon profile  for large $C$. In particular, for the  most realistic case of a mix of continuum and injected photons, this time scale is a few~million yr,  nearly independent of $C$, for large $C$.  However, this does not correspond to a situation in which both the photon profile and 
the thermal state have relaxed to steady state. For both continuum and a mix of continuum and injected photons, the time scale at which  the true  equilibrium is reached corresponds to the one given in point (a).  The  complex thermal history  that emerges from the interplay of these diverse time scales is captured in Figure~\ref{fig:cont+inj_ssint}.  

\item[(d)] Finally, we could have other  time scales owing to  the short life span of the first ionizing  sources. We show that  the source switching on and off  affects  both the profile shape and the short- and long-term thermal evolution of IGM. Recent JWST results might   have provided partial 
evidence of Population~III stars at high redshifts (e.g. \cite{hassan2023jwst,yan2022first}). Such stars have life cycles of between 0.1--10~Myr, which  are comparable to 
the time scales  we find in this paper.  
\end{itemize}

Is it possible to detect the impact of the time-dependence of Lyman-$\alpha$ heating on the \hi\ signal from CD/EoR? As Lyman-$\alpha$ photons can heat the IGM to temperatures in the range 50--100~K, which is comparable to or exceeds   the CMB temperature at $z\simeq 20$, Lyman-$\alpha$ heating could replace/supplement  X-ray heating  for large enough
Lyman-$\alpha$ photon density. This would alter both the global and fluctuating component of the \hi\ signal. In particular, given that the spatial scales affected  by X-rays and Lyman-$\alpha$ photons  are very different, Lyman-$\alpha$ heating could be detectable from the fluctuating component of the \hi\ signal from the era of CD/EoR (for details, see e.g. \cite{RS18,RS19,2006PhR...433..181F,Sethi05,2010ARA&A..48..127M,21cm_21cen} and references therein).
The time-dependence of temperature in Figure~\ref{fig:cont+inj_ssint} shows many complex features. Is it possible to detect 
some of these features in the \hi\ data from CD/EoR? To address this issue would require a more detailed analysis based on the production and build up of Lyman-$\alpha$ during the era of CD. However, our analysis allows us to partly answer this question. We show that the thermal state reaches an equilibrium within a few million yr after switching on the source, and the temperature is nearly independent of 
the photon production rate. The medium cools again after nearly 0.1~Gyr. Depending on the evolution to spin temperature, $T_{\rm S}$ (Eq.~(\ref{eq:ts_def})), the tomography of these  features might be observable  in the \hi\ signal. We note that upcoming telescope SKA1-low has the spectral resolution $\delta \nu \lesssim 10 \, \rm kHz$ \footnote{https://www.skao.int/sites/default/files/documents/d35-SKA-TEL-SKO-0000015-04\_Science\_UseCases-signed.pdf} which translates into  a time resolution of $10^4 \, \rm yr$, which is much shorter than the time scale over which the thermal profile evolves.  Therefore, in principle, most of the complex features induced by Lyman-$\alpha$ heating/cooling   (Figure~\ref{fig:cont+inj_ssint})  can be discerned from the \hi\ signal. We hope to explore this effect in detail in future work.

In this paper, we do not consider specific models of Lyman-$\alpha$ injection, based on the star-formation history of first sources. 
This could be a natural extension of our work.  Our study also does not take into account the impact of inhomogeneous  heating, which could be an important  factor in  the early evolution of  the thermal state of the gas  and therefore might have significant impact on \hi\ observables (e.g. \cite{RS19} for more details). Also, as our study provides partial evidence of how short-lived sources might  affect the 
thermal state of IGM,  we need more detailed models of such sources, in particular in the light of recent JWST results. We hope to return to these issues  in our  subsequent work. 

To sum up, we have  explored the complexity of Lyman-$\alpha$ heating/cooling in  the CD era in this paper: (a) The energy exchange between Lyman-$\alpha$ photons and neutral hydrogen  depends on  a range 
of time scales. For large  photon injection rates, the initial exchange occurs on time scales on the 
order of a million yr, leading to thermal equilibrium. For small photon injection rates, the exchange
continues for longer times with the expansion rate playing a role, (b) continuum photons only reach time-independent intensity profile after a few hundred million yr, (c) the final temperature depends mainly on the ratio of injected to continuum photons, and (d) a more complicated time-dependence of photon injection (e.g. switching sources on and off) could have a significant impact on thermal history of IGM.

\section*{Acknowledgments}
We would like to thank the anonymous referee and Shikhar Mittal for useful comments and suggestions, which helped us improve this paper.


\appendix

\section{Energy exchange between photons and atoms} \label{app:eneex}
In this appendix, we derive the energy exchange between photons and neutral atoms owing to Compton-inverse Compton scattering. We follow the treatment given in the appendix of \cite{2008cosm.book.....W} and extend it to resonant line scattering (e.g. CS06). 

Let us define incoming and outgoing frequencies and angles in both the
rest frame of the atom and the lab frame: $k, k',\alpha,\alpha'$ are
energies and angles in the rest frame with primed quantities corresponding
to final states. The corresponding quantities in the lab frame are 
$\omega, \omega',\eta,\eta'$. They are related as
\begin{eqnarray}
  k &= &\gamma \omega (-\beta \cos\eta +1), \quad {\rm and} \\
  \cos\alpha &=& {\cos\eta - \beta \over 1 -\beta \cos\eta}.
  \label{eq:lor_tra}
\end{eqnarray}
The relation between  primed quantities is the same. The inverse relations  can be obtained by replacing $\beta$ with $-\beta$ as
\begin{eqnarray}
  \omega &= &\gamma k (\beta \cos\alpha +1), \quad {\rm and} \\
  \cos\eta &=& {\cos\alpha + \beta \over 1 +\beta \cos\alpha}.
  \label{eq:inv_rel}
\end{eqnarray}
In the rest frame of the atom, the outgoing energy $k'$ is related to the
incoming energy $k$ as
\begin{equation}
  k' = {k \over 1+ k/m_p(1-\cos\theta)},
  \label{eq:comp_sca}
\end{equation}
where
\begin{equation}
\cos\theta = \cos\alpha \cos\alpha' + \cos(\phi'-\phi) \sin\alpha \sin\alpha' .
\end{equation}
Using Eq.~(\ref{eq:lor_tra}), Eq.~(\ref{eq:comp_sca}) can be expressed as
\begin{equation}
  (-\beta\gamma \cos\eta' + \gamma)\omega' = {\omega(-\beta\gamma\cos\eta+\gamma) \over 1+ k/m_p(1-\cos\theta)}.
  \label{eq:comp_sca1}
\end{equation}
Using Eq.~(\ref{eq:inv_rel}), $\eta$ and $\eta'$ can be eliminated for
angles in the rest frame. If the resulting expression is further expanded
in the first order in $k/m_p$\footnote{we retain the terms of order $\beta^2$ but
all terms of order $k\beta/m_p$ and $k^2/m_p^2$ are dropped.}, we get
\begin{equation} 
  {\omega' -\omega  \over \omega} \simeq -{k \over m_p }\left(1-\cos\theta\right) + {\beta(\cos\alpha' - \cos\alpha) \over 1+\beta \cos\alpha}.
  \label{eq:ene_exch1}
\end{equation}
Eq.~(\ref{eq:ene_exch1}) has to be  averaged over incoming and outcoming angles, electron distribution, and weighted with the cross-section.
For isotropic scattering, the outgoing direction $\alpha'$ is isotropic in the atom's rest frame while the incoming angle $\eta$ (and atom velocities) is isotropic in the lab frame. 
Notice that the result depends on the dot product of electron velocity and the photon direction, and therefore, one needs to average
over only two of these quantities.

For resonant scattering, the cross-section of the scattering $\sigma_k\propto \phi(k)$, where response function $\phi(k)$ is a sharply peaked function of 
energy $k$ in the rest frame of the atom; $\phi(k)$ is normalized  probability density of scattering, with $\int \phi(k) dk=1$. Our aim
is to compute the response to a photon whose energy is given in the lab frame. Using Eq.~(\ref{eq:lor_tra}) and expanding in small parameter $\beta$, we get
$\phi(k) = \phi(\omega) - \omega\beta\cos\eta\phi'(\omega)$. As we retain terms only upto $\beta^2$, $\cos\eta$ can be replaced by
$\cos\alpha$ to this order  in this expansion (Eq.~(\ref{eq:inv_rel})). The exchange weighted by the frequency response function
can be expressed as (to order $k/m_e$ and $\beta^2$)
\begin{eqnarray}
    \left \langle {\omega' -\omega  \over \omega}  \right \rangle_\omega &\simeq & -{k \over m_p }\left(1-\cos\theta\right) + {\beta(\cos\alpha' - \cos\alpha) \over 1+\beta \cos\alpha} 
    - {\beta^2(\cos\alpha' - \cos\alpha) \omega \phi'(\omega) \over \phi(\omega)}.
  \label{eq:ene_exch2}
\end{eqnarray}
The average over angles ($\alpha'$ and the azimuthal angle) is best carried out in the atom rest frame as it is isotropic in this frame. This gives
\begin{equation}
    \left \langle {\omega' -\omega  \over \omega}  \right \rangle_{\omega \alpha'}\simeq  -{k \over m_p } - {\beta\cos\alpha \over 1+\beta \cos\alpha} + {\beta^2 \omega \phi'(\omega)\cos^2\alpha \over \phi(\omega)}.
  \label{eq:ene_exch3}
\end{equation}
Before averaging over the incoming angle $\alpha$, it should be noted that the transition probability also depends on the relative speed
in addition to the cross-section. Taking this into account and averaging over $\alpha$ (Equations C15 and C16 of \cite{2008cosm.book.....W}), we get
\begin{equation}
    \left \langle {\omega' -\omega  \over \omega}  \right \rangle_{\omega \alpha'\alpha}\simeq  -{k \over m_p } + {4\beta^2 \over 3} + {\beta^2 \omega \phi'(\omega)\over 3\phi(\omega)}.
  \label{eq:ene_exch4}
\end{equation}
Finally, one needs to average over speeds using  Maxwellian distribution; this yields, restoring factors of $\hbar$ and $c$,
\begin{equation}
    \left \langle {\omega' -\omega  \over \omega}  \right \rangle_{\omega \alpha'\alpha v}\simeq  -{\hbar \omega \over m_p c^2 } + {4k T \over m_p c^2} + {k T \omega \phi'(\omega)\over m_p  c^2\phi(\omega)}.
  \label{eq:ene_exch5}
\end{equation}
Without the third term on RHS, this expression reduces to an energy exchange expression relevant for the derivation of Kompaneets equation. However, the crucial term in our case is the third term on the RHS,  which arises owing to resonant scattering.  It can be shown that this term dominates the second term on RHS. We drop the second term on the RHS for the rest of this paper. In this case, Eq.~(\ref{eq:ene_exch5}) reduces to the expression derived  by  CS06.

\bibliographystyle{aasjournal}

\end{document}